\documentclass[preprint,3p]{elsarticle}

\usepackage[ansinew]{inputenc}
\usepackage[english]{babel}
\usepackage{ifthen,shortvrb}
\usepackage[fleqn]{amsmath}
\usepackage{amsxtra,amsfonts,latexsym,amssymb,euscript}
\usepackage{amstext}
\usepackage{graphicx}
\usepackage{dcolumn}
\usepackage{subfigure}
\usepackage[usenames,dvipsnames]{color}
\usepackage{natbib}
\usepackage{mathrsfs}

\newcommand{\bepsilon}{\boldsymbol{\epsilon}}
\newcommand{\bsigma}{\boldsymbol{\sigma}}

\newcommand{\bu}{\mathbf{u}}
\newcommand{\bv}{\mathbf{v}}
\newcommand{\dO}{\mathrm{d}\Omega}

\newcommand{\gnn}{g_\mathrm{n}}
\newcommand{\gnnz}{g_\mathrm{n,0}}
\newcommand{\gtt}{g_\mathrm{t}}
\newcommand{\bT}{\mathbf{T}}

\begin{document}

\begin{frontmatter}

\title{A variational approach with embedded roughness for adhesive contact problems}

\author[IMT]{M.~Paggi\corref{mycorrespondingauthor}}
\cortext[mycorrespondingauthor]{Corresponding author}
\ead{marco.paggi@imtlucca.it}
\author[Seville]{J.~Reinoso}

\address[IMT]{IMT School for Advanced Studies Lucca, Piazza San
  Francesco 19, 55100 Lucca, Italy}
\address[Seville]{Elasticity and
  Grupo de Elasticidad y Resistencia de Materials, Departamento de Mec\'anica de Medios Continuos y Teor\'ia de Estructuras,  E.T.S de Ingenier\'ia, Universidad de Sevilla, Camino de los Descubrimientos s/n, 41092 Seville, Spain}

\begin{abstract}
A new variational formulation is herein proposed for the solution of adhesive contact problems for non-planar profiles of arbitrary shape indenting a deformable half-plane. The method exploits the original idea of accounting for the shape of roughness as a correction to the normal gap function, rather than explicitly discretizing roughness with higher-order numerical interpolation schemes. The resulting interface finite element with eMbedded Profile for Joint Roughness (MPJR interface finite element) is derived and its implementation as a user-defined element is comprehensively described.
The method is validated against Hertzian contact problems between a cylinder and a half-plane, also in the presence of adhesion, showing a remarkable accuracy in spite of the low-order interpolation used. The capability and efficiency of the method are subsequently illustrated in relation to the challenging frictionless normal contact problems between deformable substrates and rough profiles analytically described by the Weierstrass-Mandelbrot function. This method opens new perspectives for the solution of contact problems with roughness using the finite element method that, so far, were strongly limited by the issue of providing an accurate representation of roughness.
\end{abstract}

\begin{keyword}
Contact mechanics; roughness; waviness; adhesion; finite element method.
\end{keyword}

\end{frontmatter}


\section{Introduction}

Stress transfer between bodies separated by non-planar or rough interfaces is a timely frontier research topic in engineering and physics, as also highlighted in recent review articles \citep{mueser,vakis}. Starting from normal Hertzian contact problems with adhesion \citep{JKR,DMT,MULLER,MAUGIS,Greenwood1997,GJ,Greenwood2007}, contact mechanics has seen considerable efforts in the development of analytical theories for adhesiveless nominally flat surfaces \citep{GW66,Barber03,Persson1,Ciavarella,revitalized,carbone,CGP08,PC10,PB11}, further generalized to deal with adhesion phenomena \citep{policarpou,Persson1,G07P1,G07P2,G08}.

Numerical methods are also an important tool for the investigation of contact problems with a complex arbitrary roughness distribution, providing a way to predict not only global quantities such as the total real contact area and the total normal force vs. the applied displacement, which are typically predicted by analytical methods, but also local quantities related to the distribution of tractions, contact areas, and normal gaps across the rough surface. In this regard, the boundary element method (BEM) \citep{PK,BP15} has been proved to be very efficient from the computational viewpoint, since it requires the discretization of the interface without the need of discretizing the half-plane. However, standard BEM formulations are based on the fundamental assumptions of linear elasticity and homogeneity of the materials, and their extension to inhomogeneities \citep{nelias}, finite-size geometries \citep{finite}, or interface constitutive nonlinearities such as adhesion \citep{Rey2017,Popov2017,Popov}, are not trivial tasks.

In this regard and based on the previous discussion , the finite element method (FEM) would be conceived as  a more suitable computational approach to pursuit in order to overcome all the major limitations of BEM. This stems from the fact that FEM can easily take into account material or interface nonlinearities, and finite-size geometries due to its inherent versatility and robustness. Moreover, it is prone to be extended for the solution of multi-field problems involved in heat transfer or in reaction-diffusion systems. However, in spite of the different appealing aspects of FEM over BEM, this approach has been limited to few remarkable studies concerning contact problems in elasto-plasticity \citep{PEI}.
The motivation is primarily due to the need of discretizing the bulk and also the rough interface, which is not an easy task. As shown in \citep{HPMR,wriggers}, Bezier interpolation techniques can be employed to regularize rough interfaces to become amenable for contact search algorithms. Nevertheless, this latter  approach should  be applied with care in order to avoid artificial smoothing of finer surface features relevant for contact mechanics.

Differing from previous studies, in this work,  a new variational formulation for the solution of normal contact problems involving bodies separated by a non-planar smooth or rough interface, of any shape is originally proposed. The approach, presented in detail in Sec. 2, abandons the classical path of explicitly modeling surface roughness and, considering the interface as nominally flat, embeds the analytical expression of roughness as a correction to the normal gap. The finite element discretization associated to the weak form, detailed in Sec. 3, leads to the formulation of a new zero-thickness interface finite element with eMbedded Profile for Joint Roughness (called MPJR interface finite element) which has a straightforward implementation as a user element routine for standard implicit finite element procedures.

The validation of the proposed computational method, herein implemented in the research finite element analysis programme FEAP \citep{FEAP}, is discussed in Sec. 4 in reference to the frictionless Hertzian contact problems between a cylinder and a half-plane, with or without adhesion. The method is then fully exploited in Sec. 5 for bi-dimensional plane strain contact problems involving a profile whose multi-scale roughness is modeled according to the Weierstrass-Mandelbrot function. The case studies have been selected to test the method against some of the major complexities that pose limitations to the applicability of the state-of-the-art analytical and computational methods, namely: $(i)$ the non-compactness of the contact domain; $(ii)$ modeling multiple length scales of roughness with accuracy and with a reasonable computation cost; $(iii)$ solving contact problems with finite-size geometries; $(iv)$ accounting for highly nonlinear adhesion models based on the Lennard-Jones potential.

\section{Variational formulation with embedded roughness}

In this section, we propose the variational formulation governing the problem of contact and adhesion between two bodies across a non-planar or a rough interface. Starting from the strong differential form describing the mechanics of the continua and the problem of contact with adhesion along the interface, we derive of the corresponding weak form that provides the basis for the new interface finite element detailed in Sec. 3. Although the formulation is herein specialized to bi-dimensional domains, the proposed framework can be extended to three-dimensional topologies in a straightforward manner, exploiting the same formalism.

\subsection{Governing equations and strong form}

\begin{figure}[h!]
\begin{center}
\includegraphics[width=.6\textwidth,angle=0]{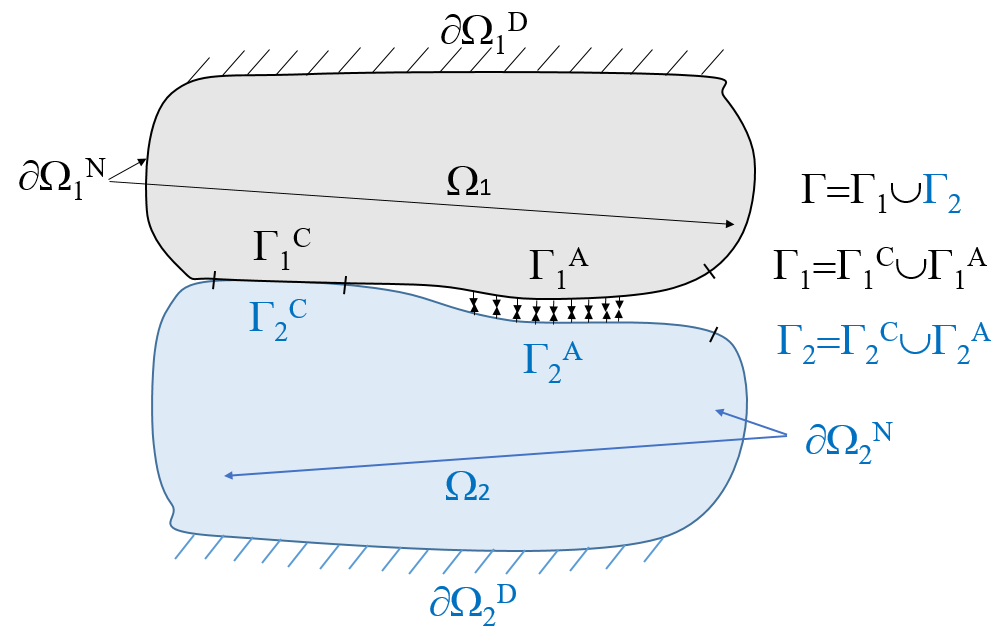}
\caption{Domains $\Omega_i$ $(i=1,2)$, their Dirichlet $(\partial\Omega_i^D)$ and Neumann $(\partial\Omega_i^N)$ boundaries, and the interface $\Gamma=\Gamma_1\bigcup \Gamma_2$ composed of an adhesive part, $\Gamma_i^A$, and a contact part, $\Gamma_i^C$.}\label{fig1}
\end{center}
\end{figure}

Let two deformable bodies occupy the domains $\Omega_i \in \mathbb{R}^2$ $(i=1,2)$ in the undeformed configuration defined by the reference system $Oxy$. The two domains are separated by an interface $\Gamma$ defined by the opposite boundaries $\Gamma_i$ $(i=1,2)$ of the two bodies, viz. $\Gamma=\bigcup_{i=1,2} \Gamma_i$, where contact or adhesive interactions take place. The whole boundary of the $i$-th body, $\partial \Omega_i$, is therefore splitted into three parts: $(i)$ a portion where displacements are imposed, i.e., the Dirichlet boundary $\partial \Omega_i^D$; $(ii)$ a portion where tractions are specified, i.e., the Neumann boundary $\partial \Omega_i^N$; $(iii)$ and the interface $\Gamma_i=\Gamma_i^C\bigcup \Gamma_i^A$ where specific boundary conditions have to be imposed to model contact on $\Gamma_i^C$ or adhesion on $\Gamma_i^A$, see Fig.\ref{fig1}. The partition of $\Gamma_i$ in $\Gamma_i^C$ and $\Gamma_i^A$ is not known a priori, but it is the result of the solution of the elastic problem.

\begin{figure}[h!]
\begin{center}
\includegraphics[width=.55\textwidth,angle=0]{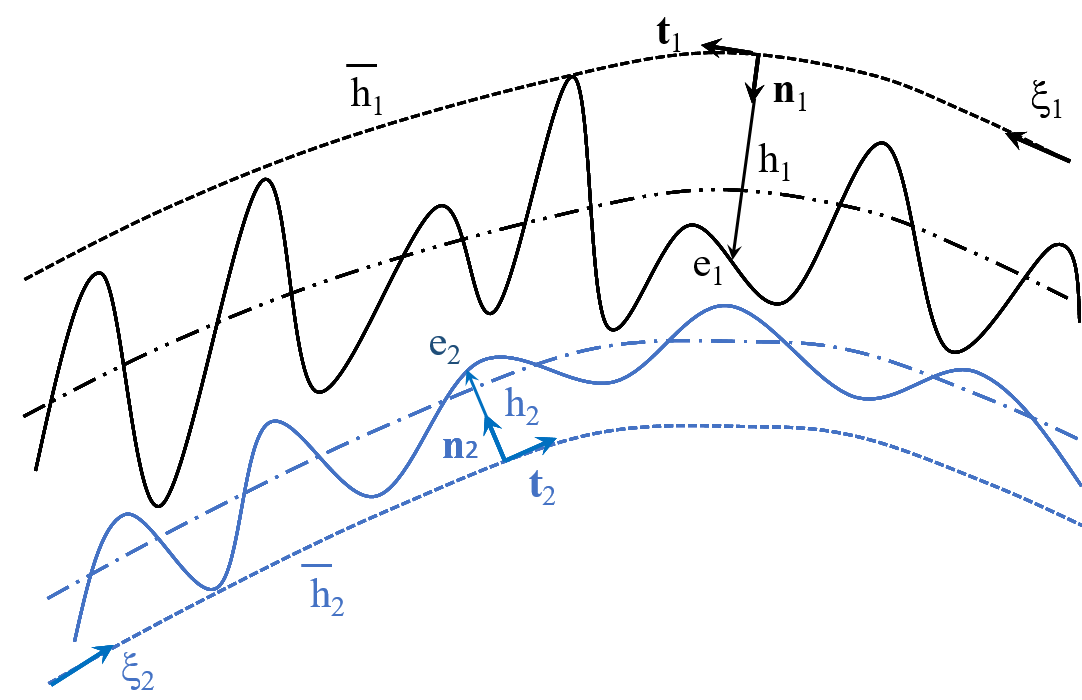}
\caption{Parametrization of two microscopically rough profiles composing an interface $\Gamma$.}\label{fig2}
\end{center}
\end{figure}

Here, we assume that $\Gamma_i$ are \emph{nominally flat but microscopically embedding rough profiles}. $\Gamma_1$ and $\Gamma_2$ can be one the negative of the other, as in the case of an interface originated by fracture, or different from each other, as for two bodies coming into contact, without any restriction.
We now introduce for the $i$-th rough profile $\Gamma_i$ its \emph{smoother representation} $\overline{h}_i(\xi_i)$, given by a curve parallel to the average line of the profile and with datum set in correspondence of its deepest valley (see Fig.\ref{fig2}). A point along the curve $\overline{h}_i(\xi_i)$ is identified by a value of the curvilinear coordinate $\xi_i=\xi_i(x,y)$, which establishes a one-to-one correspondence with the coordinates of the same point in the global reference system $Oxy$. We also associate the tangential and normal unit vectors $\mathbf{t}_i(\xi_i)$ and $\mathbf{n}_i(\xi_i)$ to $\overline{h}_i(\xi_i)$, to identify the normal and the tangential directions at any point along the smoothed curve $\overline{h}_i(\xi_i)$, with $\mathbf{n}_i$ pointing outwards from the domain $\Omega_i$. The actual elevation of the rough profile measured from $\overline{h}_i(\xi_i)$ is finally described by the \emph{roughness function} $h_i(\xi_i)$. Therefore, the $i$-th boundary $\Gamma_i$ is herein parametrized such that its actual elevation $e_i(\xi_i)$ in the curvilinear setting is given by $e_i(\xi_i)=\overline{h}_i(\xi_i)+h_i(\xi_i)$.

\begin{figure}[h!]
\begin{center}
\subfigure[Composite topography of the interface displaying a rough profile]{\includegraphics[width=.6\textwidth,angle=0]{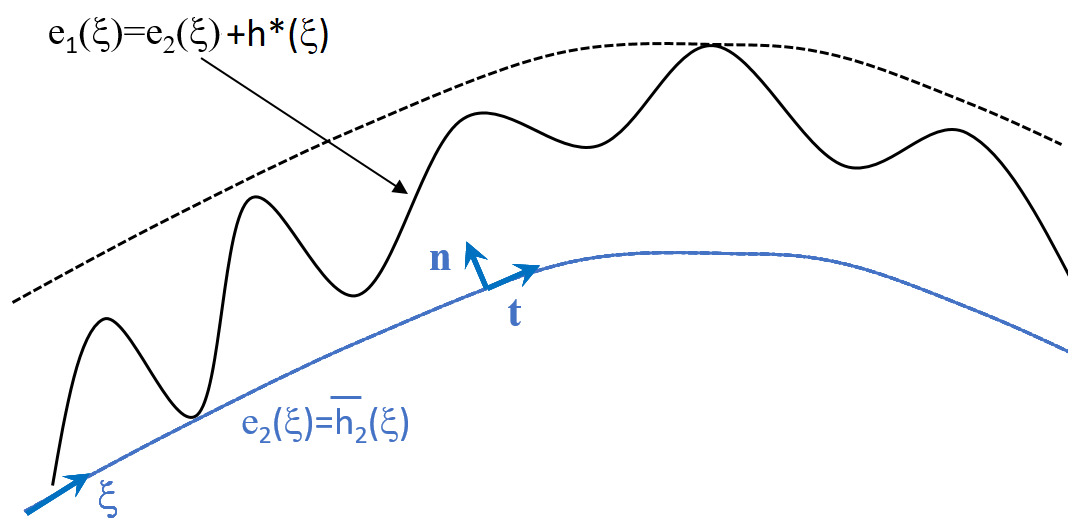}\label{fig3a}}\\
\subfigure[Zero-thickness interface model]{\includegraphics[width=.6\textwidth,angle=0]{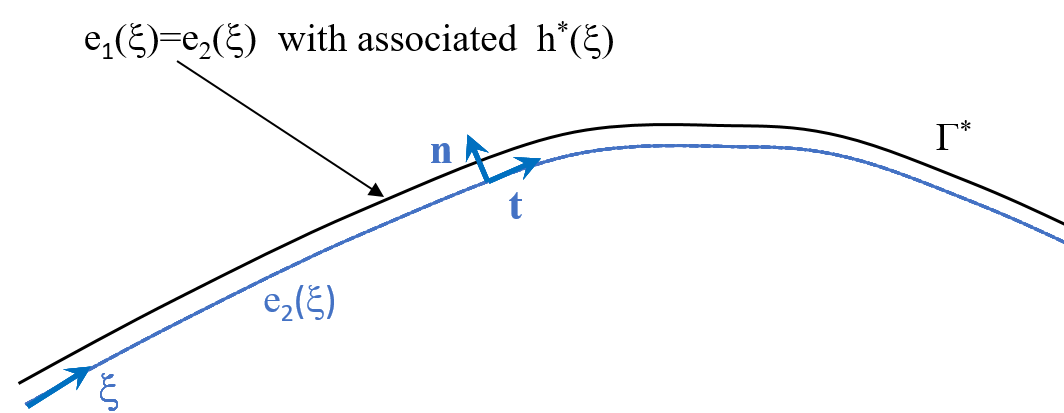}\label{fig3b}}
\caption{Composite topography of the interface $\Gamma$ (a), and its zero-thickness interface representation $\Gamma^*$ (b). }\label{fig3}
\end{center}
\end{figure}

To simplify the matter, we now introduce the \emph{composite topography} $\Gamma^*$ of the interface $\Gamma$, represented by a flat curve, $e_2=\overline{h}_2(\xi)$, and a profile with elevation $e_1(\xi)=\overline{h}_2(\xi)+h^*(\xi)$, where $h^*(\xi)=\max_{\xi}[h_1(\xi)+h_2(\xi)]-[h_1(\xi)+h_2(\xi)]$. After this transformation, a \emph{zero-thickness interface model} for $\Gamma^*$ is introduced and defined by the two initially coincident but distinct (not-joined) flat curves described by the function $e_2(\xi)$, plus the associated function $h^*(\xi)$. This composite topography has also unique tangential and normal unit vectors $\mathbf{t}$ and $\mathbf{n}$, conventionally associated to the tangential and normal unit vectors computed at any point of the flat curve $e_2$, see Fig.\ref{fig3}.

We now postulate the existence of a displacement field for each body, $\bu_i=(u_i,v_i)^\text{T}$, that maps the transformation from the underformed configuration to the deformed one, and viceversa. Such functions are thereby assumed to be continuous, invertible and differentiable functions of the position vector $\mathbf{x}=(x,y)^{\text{T}}$ within each body. At the interface $\Gamma^*$, on the other hand, the configuration of the system is described by the relative displacement field $\Delta\bu$, usually denominated as gap field across the interface $\mathbf{g}$,  which is mathematically defined as  the projection of the relative displacement $\bu_1-\bu_2$ onto the normal and tangential directions of the interface defined by the unit vectors $\mathbf{n}$ and $\mathbf{t}$, respectively. In components, the vector $\Delta\bu$ collects the relative tangential displacement, $\Delta u_t$, and the relative normal displacement, $\Delta u_n$, i.e., $\Delta\bu=(\Delta u_t,\Delta u_n)^{\text{T}}$.

In the present framework, the normal gap $\gnn$ of the composite topography, which represents the actual physical separation between the composite topography and the smooth curve $e_2$ after deformation, is given by $\gnn=\Delta u_n+h^*$. Based on the value of $\gnn$, the portion of the interface in contact, $\Gamma^C=\Gamma_1^C\bigcup \Gamma_2^C$, is identified by the condition $\gnn=0$. On the other hand, the portion subject to adhesion, $\Gamma_A=\Gamma_1^A\bigcup \Gamma_2^A$, presents a positive-valued normal gap $\gnn>0$. A negative valued normal gap is not admissible so far by definition, since it would imply  the compenetration between the bodies.

Inside each deformable material, the small deformation strain tensor $\bepsilon_i$ $(i=1,2)$ is introduced as customary, which is defined as the symmetric part of the   displacement gradient: $\bepsilon_i =  \nabla^{s} \bu_i$. In the sequel, the standard Voigt notation will be used and the strain tensor components will be collected in the vector  $\bepsilon_i=(\epsilon_{xx},\epsilon_{yy},\gamma_{xy})_i^{\text{T}}$.

In the absence of body forces, the strong (differential) form of equilibrium for each body is provided by the linear momentum equation along with the Dirichlet and the Neumann boundary conditions on $\partial \Omega_i^D$ and $\partial \Omega_i^N$, respectively $(i=1,2)$, equipped by the conditions for contact on $\Gamma^*_C$ and adhesion on $\Gamma^*_A$:
\begin{subequations}\label{strong}
\begin{align}
\nabla \cdot \bsigma_i &=\mathbf{0} \;\; \text{in}\, \Omega_i,\\
\bu_i &=\overline{\bu} \;\; \text{on}\, \partial\Omega_i^D,\\
\bsigma_i \cdot \mathbf{n} &= \mathbf{T}\;\; \text{on}\, \partial\Omega_i^N, \\
\gnn &= 0,\, p_n<0\;\; \text{on}\, \Gamma^*_C\\
\gnn &> 0,\, p_n=p_A>0\;\; \text{on}\, \Gamma^*_A,
\end{align}
\end{subequations}
where $\overline{\bu}$ denotes the imposed displacement, $\mathbf{T}$ the applied traction vector, and $p_A(g_n)$ is a function of the relative displacement $\Delta\bu$. Therefore, the nonlinearity of the problem stems from the fact that the contact and adhesive portions of the interface $\Gamma^*$ are known only once the displacement field, solution of the problem, is known. As a consequence, the present problem can be ascribed to the family of the so-called moving boundary value problems and it requires an iterative solution scheme.

For its solution, the strong form has to be equipped by the constitutive equations for the bulk and for the interface. For the bulk, recalling standard thermodynamic arguments, general  (linear or nonlinear) constitutive stress-strain relations can be postulated without any loss of generality for the $i$-th material domain:  $\bsigma_i := \partial_{\bepsilon_i} \Psi (\bepsilon_i)$ and 
$\mathbb{C}_i : = \partial^{2}_{\bepsilon_i \bepsilon_i} \Psi (\bepsilon_i)$, whereby $\Psi (\bepsilon_i)$ is the Helmholtz free-energy function for the  body $i$, whereas its corresponding Cauchy stress tensor and the constitutive operator are respectively denoted by  $\bsigma_i$ and $\mathbb{C}_i$. The two bodies are in general both deformable, but in the present setting it is also possible to consider one of them as rigid. This condition is of paramount interest for contact mechanics in the presence of two dissimilar linear elastic bodies. In such a case, it is possible to simplify the matter by replacing the bi-material system by a rigid body indenting a linear elastic material having a composite Young's modulus $E=[(1-\nu_1)/E_1+(1-\nu_2)/E_2]^{-1}$, function of the Young's moduli $E_i$ and Poisson's ratios $\nu_i$ $(i=1,2)$ of the two elastic materials, see \citep{barber_contact,barber_elasticity}.

Regarding the interface, the constitutive response should be introduced by distinguishing between the normal and the tangential directions. Although the present formulation can encompass any type of loading condition, we restrict our attention in this study to the frictionless normal contact problem with or without adhesion. In general, the constitutive relation in the tangential direction should account for frictional effects and adhesion, and it is left for further investigation.

In the normal direction, the contact condition imposes that the displacement field solution leads to a vanishing normal gap $\gnn=0$. Correspondingly, contact tractions are negative valued. On the other hand, for $\gnn>0$, positive-valued adhesive tractions apply and are herein modeled by a relation dictated by an adhesion model inspired by the Lennard-Jones potential, see \citep{policarpou} for its use in contact mechanics. In formulae, we use the following expression for $\gnn>0$:
\begin{equation}\label{L-J}
p_A=24\varepsilon\left[\dfrac{\kappa^6}{(\gnn+\gnnz)^7}-2\dfrac{\kappa^{12}}{(\gnn+\gnnz)^{13}}\right]
\end{equation}
where $\varepsilon$ and $\kappa$ are model parameters and $\gnnz$ is the molecular equilibrium distance. Clearly, this adhesive constitutive relation represents another source of nonlinearity. For vanishing adhesive tractions $(\varepsilon=0)$, the present model leads to the classical unilateral contact condition as a limit case.

\subsection{Weak form}

According to the principle of virtual work, the weak form associated with the strong form Eq.\eqref{strong} reads:
\begin{equation}\label{weak}
\begin{aligned}
\Pi=&\int_{\Omega_1}\bsigma_1(\bu_1)^{\text{T}}\bepsilon_1(\bv_1)\dO+\int_{\Omega_2}\sigma_2(\bu_2)^{\text{T}}\bepsilon_2(\bv_2)\dO
-\int_{\partial\Omega_1^N}\bT^{\text{T}} \bv_1\text{d}\partial\Omega
-\int_{\partial\Omega_2^N}\bT^{\text{T}} \bv_2\text{d}\partial\Omega\\
&-\int_{\Gamma^*_C}p_n(\Delta\bu) g_n(\Delta\bv)\text{d}\Gamma-\int_{\Gamma^*_A}p_A(\Delta\bu) g_n(\Delta\bv)\text{d}\Gamma=0,
\end{aligned}
\end{equation}
where $\bv_i$ is the test function (virtual displacement field) and $g_n(\Delta\bv)$ is the virtual normal gap at the interface $\Gamma^*$. The test function in the $i$-th body fulfills the condition $\bv_i=\mathbf{0}$ on $\partial\Omega_i^D$ and the adhesive-contact condition on  $\Gamma^*$. The adhesive-contact conditions on $\Gamma^*$ impose that the corresponding integrals are greater or equal to zero everywhere on $\Gamma^*$. Thus, the solution of the problem implies the solution of the following variational inequality:
\begin{equation}
\int_{\Omega_1}\bsigma_1(\bu_1)^{\text{T}}\bepsilon_1(\bv_1)\dO+\int_{\Omega_2}\sigma_2(\bu_2)^{\text{T}}\bepsilon_2(\bv_2)\dO
-\int_{\partial\Omega_1^N}\bT^{\text{T}} \bv_1\text{d}\partial\Omega
-\int_{\partial\Omega_2^N}\bT^{\text{T}} \bv_2\text{d}\partial\Omega\ge 0.
\end{equation}

For the treatment of the adhesive-contact conditions, we propose to relax the constraint on the sign of the normal gap $g_n$ by using a generalized penalty approach able to deal with both contact and adhesion, see also \citep{Paggi1,Paggi2} in the context of contact and cohesive fracture at interfaces. The modified weak form \eqref{weak} reads:
\begin{equation}\label{weak-penalty}
\begin{aligned}
\Pi=&\int_{\Omega_1}\bsigma_1(\bu_1)^{\text{T}}\bepsilon_1(\bv_1)\dO+\int_{\Omega_2}\sigma_2(\bu_2)^{\text{T}}\bepsilon_2(\bv_2)\dO
-\int_{\partial\Omega_1^N}\bT^{\text{T}} \bv_1\text{d}\partial\Omega
-\int_{\partial\Omega_2^N}\bT^{\text{T}} \bv_2\text{d}\partial\Omega\\
&-\int_{\Gamma^*}p_n(\Delta\bu) g_n(\Delta\bv)\text{d}\Gamma
\end{aligned}
\end{equation}
where $p_n$ is given by:
\begin{equation}\label{pn}
p_n(g_n)=
\begin{cases}
K g_n, & \text{if}\; g_n<0,\\
p_A, & \text{if}\; g_n\ge 0.
\end{cases}
\end{equation}
The previous  parameter $K$ stands for  the so-called penalty stiffness, which should be  chosen high enough to avoid material penetration between adjacent bulk bodies, while $p_A$ is the adhesive traction whose expression is given in Eq.$\eqref{L-J}$. It is worth remarking here that the body is deformable in tension and in compression, which is a situation that can lead to algebraic and numerical challenges due to the strong nonlinearity of the Lennard-Jones force law. As discussed in \citep{barber_contact} [Sec. 12.4.2], tractions are very sensitive to changes in the gap in the compressive range, which is not the case in tension for sufficient large gaps. Therefore, iterative schemes optimized to handle compression or tension separately from each other are ill-suited and a rigid response in compression is usually assumed to simplify this matter.

The displacement field $\bu_i$ solution of the weak form Eq.\eqref{weak-penalty} is such that it corresponds to the minimum of $\Pi$ for any choice of the test functions $\bv_i$.

\section{Proposed interface finite element with eMbedded Profile for Joint Roughness (MPJR interface finite element)}

The numerical treatment of the penalized weak form given in Eq.\eqref{weak-penalty} within the finite element method requires the introduction of two different types of finite element discretization, one for the bulk, $\Omega_{i,h}$, and another for the interface, $\Gamma^*_{h}$, where the subscript $h$ refers to the respective discretized geometrical feature. For the bulk, standard linear quadrilateral or triangular isoparametric finite elements can be invoked, see classical finite element textbooks \citep{FEAP}.

At the interface, we assume to have conforming  finite element discretizations for the continua in terms of the interpolation of the displacement field and the spatial discretization scheme. Consequently,  we can introduce a special interface finite element with eMbedded Profile for Joint Roughness (MPJR interface finite element) whose kinematics departs from the formulation of interface elements used in nonlinear fracture mechanics for cohesive crack growth \citep{OP99,PW11b,PW12,Reinoso2014,Paggi2015}, and it is further specialized for the present contact/adhesive problem with a non-planar interface.

\begin{figure}[h!]
\begin{center}
\includegraphics[width=.5\textwidth,angle=0]{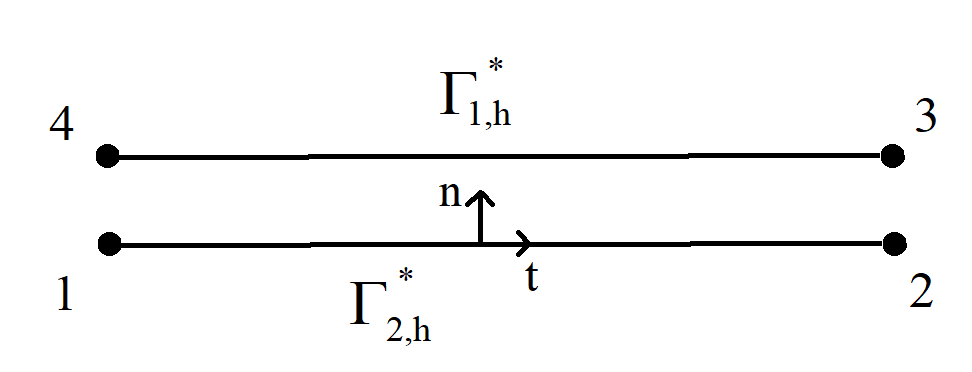}
\caption{Sketch of the interface finite element topology.}\label{fig4}
\end{center}
\end{figure}

Based on that, the interface element is defined by nodes $1$ and $2$, which belong to $\Gamma^*_{2,h}$, and by nodes $3$ and $4$, which belong to $\Gamma^*_{1,h}$, see Fig.\ref{fig4}.

The contribution of the interface to the weak form is provided by the integral $\int_{\Gamma^*}p(g_n) g_n\text{d}\Gamma$ in Eq.\eqref{weak-penalty}, which can be computed as the sum of the contributions of the whole interface elements, invoking the property of compactness of isoparametric shape functions:
\begin{equation}
\int_{\Gamma^*}p(g_n) g_n\text{d}\Gamma\cong \int_{\Gamma_h^*}p(g_n) g_n\text{d}\Gamma=\mathcal{A}_{e=1}^{n_{el}}\left\{\int_{\Gamma_e^*}p(g_n) g_n\text{d}\Gamma\right\},
\end{equation}
where the subscript $e$ refers to the $e$-th interface element $e=1,\dots,n_{el}$, and $\mathcal{A}$ symbolically denotes an assembly operator.

The interface integral is herein computed exactly, by using the 2 points Newton-Coates quadrature formula which implies the sampling of the integrand at the nodes $1$ and $2$ (or, equivalently, at nodes $3$ and $4$):
\begin{equation}
\int_{\Gamma_e^*}p(g_n) g_n\text{d}\Gamma=\sum_{j=1,2}p_i(\gnn) g_{n,i}\det J,
\end{equation}
where $\det J$ is the standard determinant of the Jacobian of the transformation that maps the geometry of the interface element from its global reference frame to the natural reference system.

To evaluate the normal gap $g_n$ at any point inside the interface element, we need to introduce the nodal displacement vector $\mathbf{d}=(u_1,v_1,\dots,u_4,v_4)^{\text{T}}$, which collects the displacements $u$ and $v$ of the four interface finite element nodes. The relative displacement $\Delta\bu$ for the nodes $1$-$4$ and $2$-$3$ is then computed by applying a matrix operator $\mathbf{L}$ which makes the difference between the displacements of nodes $1$ and $4$ with respect to those of nodes $2$ and $3$. The relative displacement within the interface finite element is then given by the linear interpolation of the corresponding nodal values, performed by the multiplication with the matrix $\mathbf{N}$ which collects the shape functions at the element level. Finally, the tangential and the normal gaps are determined by the multiplication with the rotation matrix $\mathbf{R}$ defined by the components of the unit vectors $\mathbf{t}$ and $\mathbf{n}$. In formulae, we have:
\begin{equation}
\mathbf{\Delta\bu}=\mathbf{R}\mathbf{N}\mathbf{L}\mathbf{d},
\end{equation}
where the operators present the following matrix form:
\begin{subequations}
\begin{align}
\mathbf{L} &=
\left[
\begin{matrix}
-1 &  0 &  0  &  0 &  0 & 0  & 1 &  0\\
 0 & -1 &  0  &  0 &  0 & 0  & 0 &  1\\
 0 &  0 & -1  &  0 &  1 & 0  & 0 &  0\\
 0 &  0 &  0  & -1 &  0 & 1  & 0 &  0
\end{matrix}
\right],\\
\mathbf{N} &=
\left[
\begin{matrix}
N_1 &  0    &  N_2  &  0 \\
    0 & N_1 &  0      &  N_2
\end{matrix}
\right],\\
\mathbf{R} &=
\left[
\begin{matrix}
t_x & t_y\\
n_x   & n_y
\end{matrix}
\right],
\end{align}
\end{subequations}
where $n_x$, $n_y$, $t_x$ and $t_y$ are the components of the unit vectors $\mathbf{n}$ and $\mathbf{n}$ along the $x$ and $y$ directions.

Once $\Delta\bu=(\Delta u_t,\Delta u_n)^{\text{T}}$ is determined, the actual normal gap is given by a correction to $\Delta u_n$ to account for the embedded profile that models the non-planarity of $\Gamma^*$, i.e., $\gnn=\Delta u_n+h^*$. The normal gap is used to compute the normal traction $p_n$ according to Eq.\eqref{pn}. Similarly, for further extensions to adhesive-contact problems with friction in the tangential direction, a relationship between the shearing traction $p_t$ and the relative sliding displacement $\gtt=\Delta u_t$, or its velocity, should be introduced, in analogy with the normal problem.

Due to the intrinsic nonlinearity, a full Newton-Raphson iterative-incremental scheme is herein adopted to solve the implicit nonlinear algebraic system of equations resulting from the finite element discretization:
\begin{subequations}\label{N-R}
\begin{align}
\mathbf{K}^{(k)} \Delta\mathbf{d}^{(k)} &= -\mathbf{R}^{(k)},\\
\mathbf{d}^{(k+1)} &=\mathbf{d}^{(k)}+\Delta\mathbf{d}^{(k)},
\end{align}
\end{subequations}
where the superscript $k$ denotes the iteration inside the Newton-Raphson loop, and the residual vector $\mathbf{R}_{e}^{(k)}$ and the tangent stiffness matrix $\mathbf{K}_{e}^{(k)}$ associated with the $e-$th interface finite element, assembled to the global residual vector $\mathbf{R}$ and global stiffness matrix $\mathbf{K}$ are:
\begin{subequations}\label{R-K}
\begin{align}
\mathbf{R}_{e}^{(k)} &=\int_{\Gamma^*_e} \mathbf{L}^{\text{T}}\mathbf{N}^{\text{T}}\mathbf{R}^{\text{T}}\mathbf{p}\,\text{d}\Gamma,\\
\mathbf{K}_{e}^{(k)} &=\int_{\Gamma^*_e} \mathbf{L}^{\text{T}}\mathbf{N}^{\text{T}}\mathbf{R}^{\text{T}}\mathbb{C}\mathbf{R}\mathbf{N}\mathbf{L}\,\text{d}\Gamma,
\end{align}
\end{subequations}
where $\mathbf{p}=(p_t,p_n)^{\text{T}}=(0,p_n)^{\text{T}}$ and $\mathbb{C}$ is the linearized interface constitutive matrix:
\begin{equation}
\mathbb{C} =
\left[
\begin{matrix}
\dfrac{\partial p_t}{\partial \gtt} &  \dfrac{\partial p_t}{\partial \gnn}\\
\dfrac{\partial p_n}{\partial \gtt} &  \dfrac{\partial p_n}{\partial \gnn}
\end{matrix}
\right],
\end{equation}
where, for the frictionless normal contact problem, we just need to specify $\partial p_n/\partial\gnn$ depending on the sign of the normal gap, distinguishing between the penalty relation in compression or the adhesive relation in tension.

\section{Model validation}

\subsection{Adhesiveless Hertzian contact problem between a cylinder and a half plane}

As a benchmark problem to demonstrate the capabilities of the present variational formulation with embedded roughness, we simulate the bi-dimensional frictionless normal contact problem without adhesion between a rigid cylinder indenting a half plane. For comparison purposes, we  recall the Hertzian analytical solution,  which  is available to assess the model accuracy.

\begin{figure}[h!]
\begin{center}
\includegraphics[width=1.0\textwidth,angle=0]{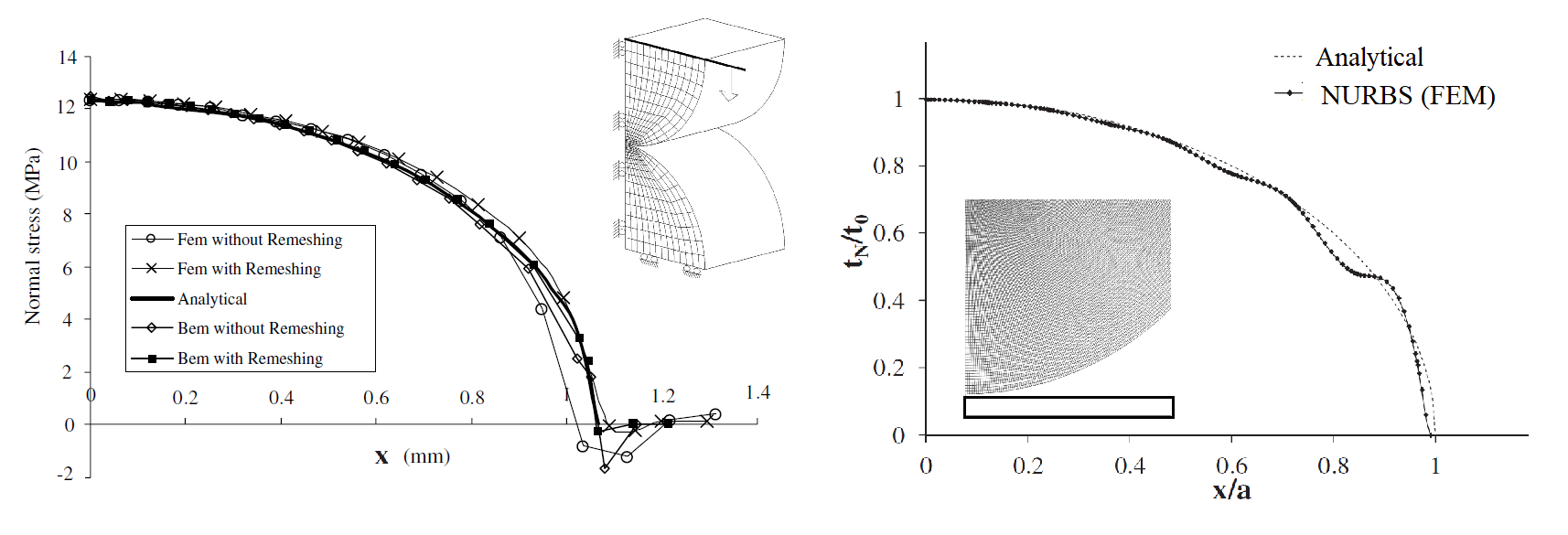}
\caption{Problems in resolving contact tractions in Hertzian normal contact problems arising from: BEM and FEM with or without mesh refinement  (figure on the left adapted from \citep{oysu}), and NURBS FEM (figure on the right adapted from \citep{laura}).}\label{fig5}
\end{center}
\end{figure}

The standard procedure for solving this problem within the finite element method requires modelling of the circular cross-section of the cylinder and the use of a contact formulation to enforce the unilateral contact constraint along the interface between the cylinder and the half plane. For that, among the possible numerical strategies, the penalty approach, the Lagrange multiplier method, and the mortar method are among the most popular formulations, see \citep{wriggers_book}. In spite of the simplicity of this non-conforming contact problem, it is well-known that all such methods require very fine meshes to resolve the contact area and the contact traction distribution, especially near the edges of the contact strip. This is primarily due to the fact that a $C^1$ linear finite element interpolation scheme is not sufficiently accurate to describe the circular shape of the cylinder. To overcome this drawback and increase the accuracy in BEM and FEM, adaptive mesh refinement has been proposed in \citep{oysu}, see Fig.\ref{fig5}(a). Alternatively, the NURBS finite element technique, which adopts shape functions with a very high regularity and smoothness, has been demonstrated to provide the best accuracy over other discretization techniques, in spite of the fact a fine mesh is in any case still required (see Fig.\ref{fig5}(b) for an example presented in \citep{laura} for the frictionless normal contact problem between a sphere and a half plane with a mesh size finer than that used within the present formulation and still showing some difficulties in capturing the analytical results).

\begin{figure}[h!]
\begin{center}
\subfigure[Actual geometry]{\includegraphics[width=.4\textwidth,angle=0]{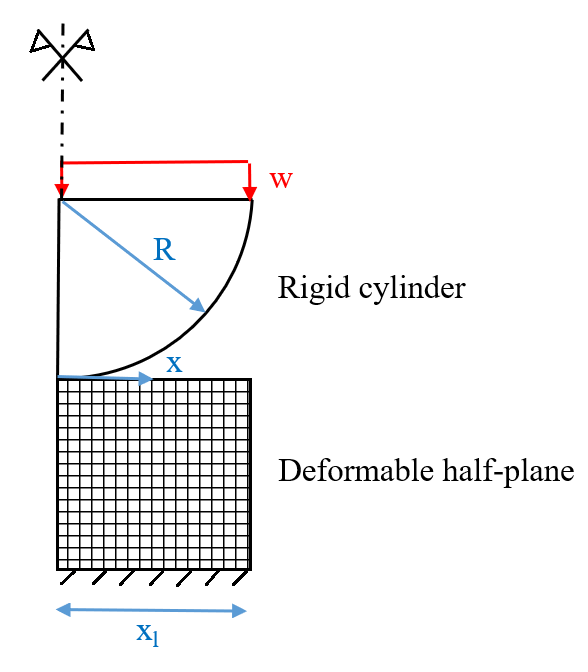}\label{fig6a}}\quad\quad 
\subfigure[Present model]{\includegraphics[width=.4\textwidth,angle=0]{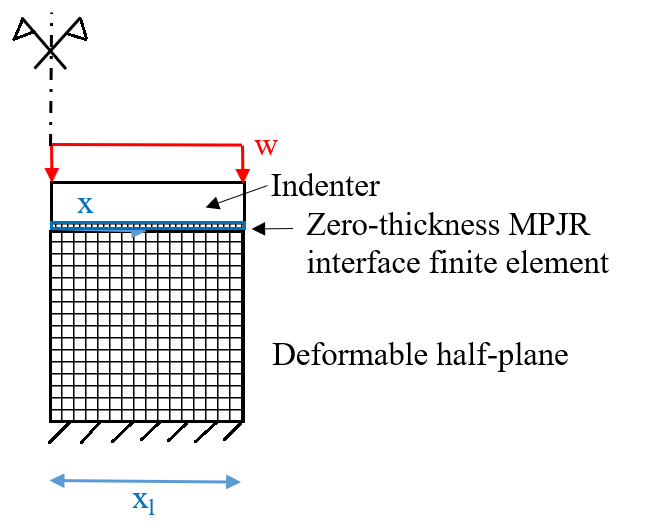}\label{fig6b}}
\caption{The actual geometry of the Hertz contact problem (a), and its finite element model based on the present variational approach with embedded roughness, which incorporates the analytical expression of the curved interface profile into the MPJR interface finite elements instead of discretizing the non-planarity of the interface.}\label{fig6}
\end{center}
\end{figure}

In the present framework, instead of modeling the geometry of the circular cross-section, the non-planarity of the interface is simply embedded in the interface element with its \emph{exact analytical function}. The actual circular shape of the boundary $\Gamma_1$ is therefore given by the composite topography of the interface profile: $e_1(x)=\overline{h}_2+h^*(x)$, where $\overline{h}_2=x_l$ and $h^*(x)=R-\sqrt{R^2-x^2}$. This strategy resolves the issues related to the accuracy of finite element interpolation schemes for the bulk and the interface, which can now have low-order linear shape functions. In this context, the geometry of the cylinder of radius $R$ occupying the domain $\Omega_1$ is replaced by a rectangular block of lateral size $x_l$ and thickness $x_l/20$, while the half-plane occupying the domain $\Omega_2$ is modeled as a plane strain domain with size $x_l$,
see Fig.\ref{fig6}.

To achieve the condition of a rigid cylinder pressed onto an elastic deformable half-plane, we set  $E_1=1000 E_2$, where the subscripts 1 and 2 identify the rigid (indenter)  and deformable (half-plane) bodies, respectively. We also set $\nu_1=\nu_2=0$ to avoid coupling between the normal and the tangential contact problems, since we have not specified a frictional constitutive response for the interface. Therefore, the composite elastic parameters are $E\cong E_2$ and $\nu=0$.

Regarding the ratio between the cylinder radius and the lateral size of the half plane, $R/x_l$, we examine two cases: $(i)$ $R/x_l=100$, which corresponds to a slightly non planar interface; $(ii)$ $R/x_l=1$, which corresponds to a significant deviation of the interface from the non-planarity. For both cases, uniform meshes for the domains $\Omega_1$ and $\Omega_2$ are used, employing four nodes linear finite elements for the bulk and the proposed MPJR interface elements with embedded roughness for the interface. The whole interface is discretized in the horizontal direction by only $n_{el}=100$ finite elements, which is much less than what is used in NURBS to achieve the same accuracy, and without adopting any mesh refinement.

Dirichlet boundary conditions are represented by imposed downwards vertical displacements $w$ on the topmost side of the domain $\Omega_1$, monotonically increasing with a pseudo-time variable to simulate the quasi-static normal contact problem; a fully restrained lower side of the domain $\Omega_2$; and a symmetry condition on the vertical size of domains $\Omega_1$ and $\Omega_2$ to account for the symmetry in the geometry and in the loading (Fig.~\ref{fig6b}).

Numerical predictions are provided in terms of a dimensionless normal contact pressure, $p/E$ vs. the dimensionless position along the interface, $x/R$. The contact pressure $p$ is given by $p=-p_n$, and therefore it is positive valued on the portion of the interface $\Gamma_C$ in contact and it must be zero elsewhere, since adhesion is not considered here. The penalty stiffness $K$ is set $K=10 E_1/x_l$, to model a very stiff interface and avoid material penetration.

\begin{figure}[h!]
\begin{center}
\subfigure[$R/x_l=100$]{\includegraphics[width=.75\textwidth,angle=0]{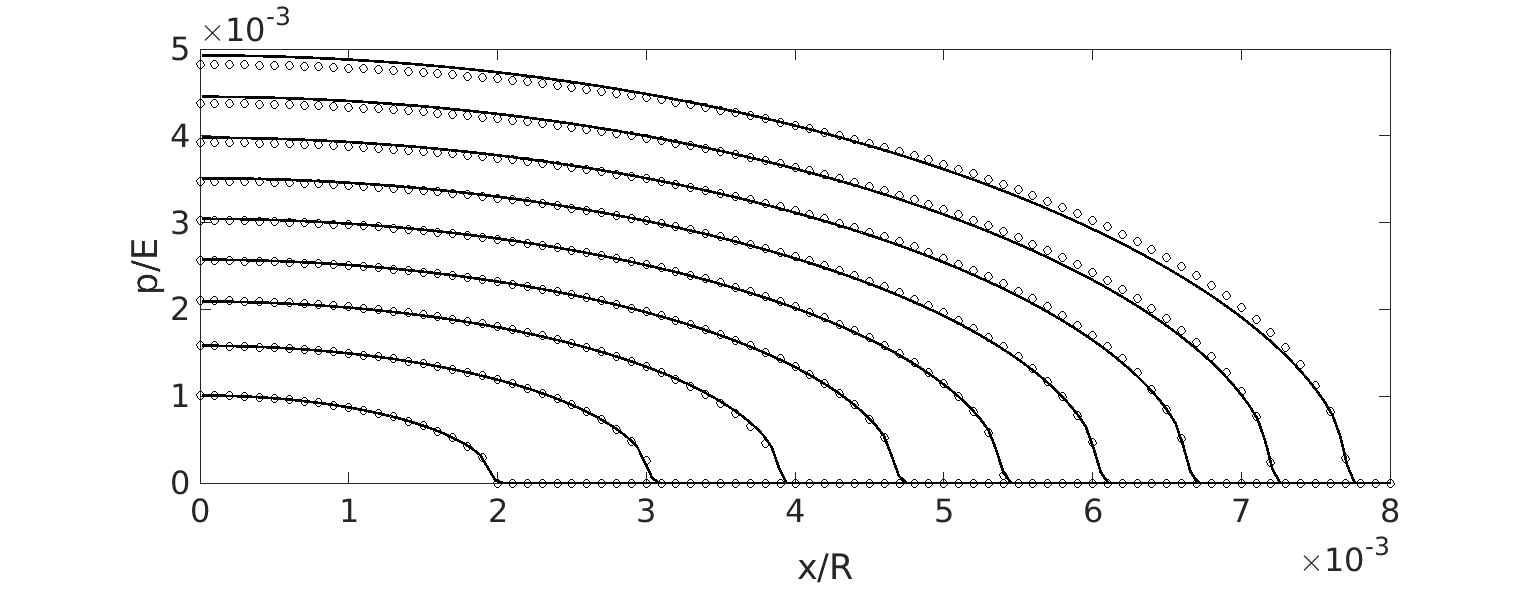}\label{fig7a}}\\
\subfigure[$R/x_l=1$]{\includegraphics[width=.75\textwidth,angle=0]{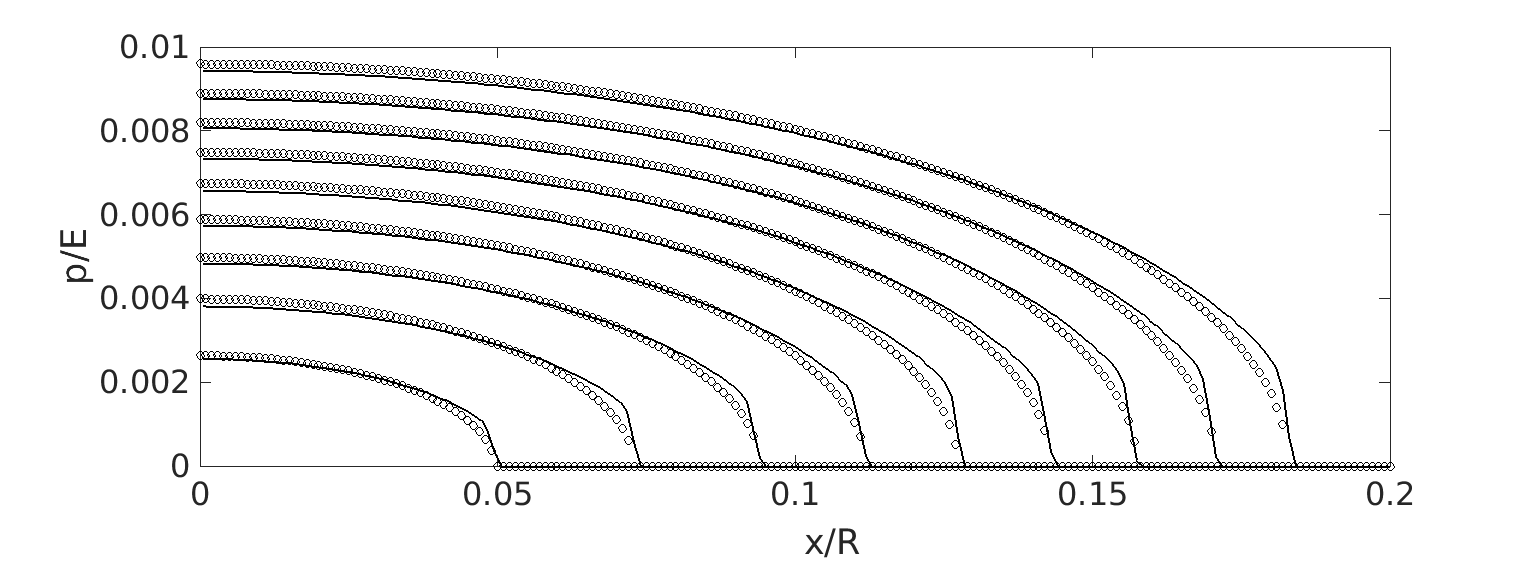}\label{fig7b}}
\caption{Dimensionless contact pressure along the interface for different imposed far field displacements. $E$, $R$ and $x_l$ denote, respectively, the composite Young's modulus, the cylinder radius, and the lateral size of the domain, respectively. The analytical Hertzian solution is superimposed with circles.}\label{fig7}
\end{center}
\end{figure}

Results from the current simulations are shown in Fig.\ref{fig7a} for the case $R/x_l=100$ and in Fig.\ref{fig7b} for $R/x_l=1$, considering 9 increasing values for the imposed far-field displacement $w$. Analytical Hertzian results, corresponding to the same contact radii, are also superimposed by circles. As can be observed in these graphs, the agreement between the present model predictions and theory is excellent, also for the case $R/x_l=1$, which is indeed very challenging from the computational point of view due to the significant non-planarity of the interface.

\begin{figure}[h!]
\begin{center}
\includegraphics[width=.7\textwidth,angle=0]{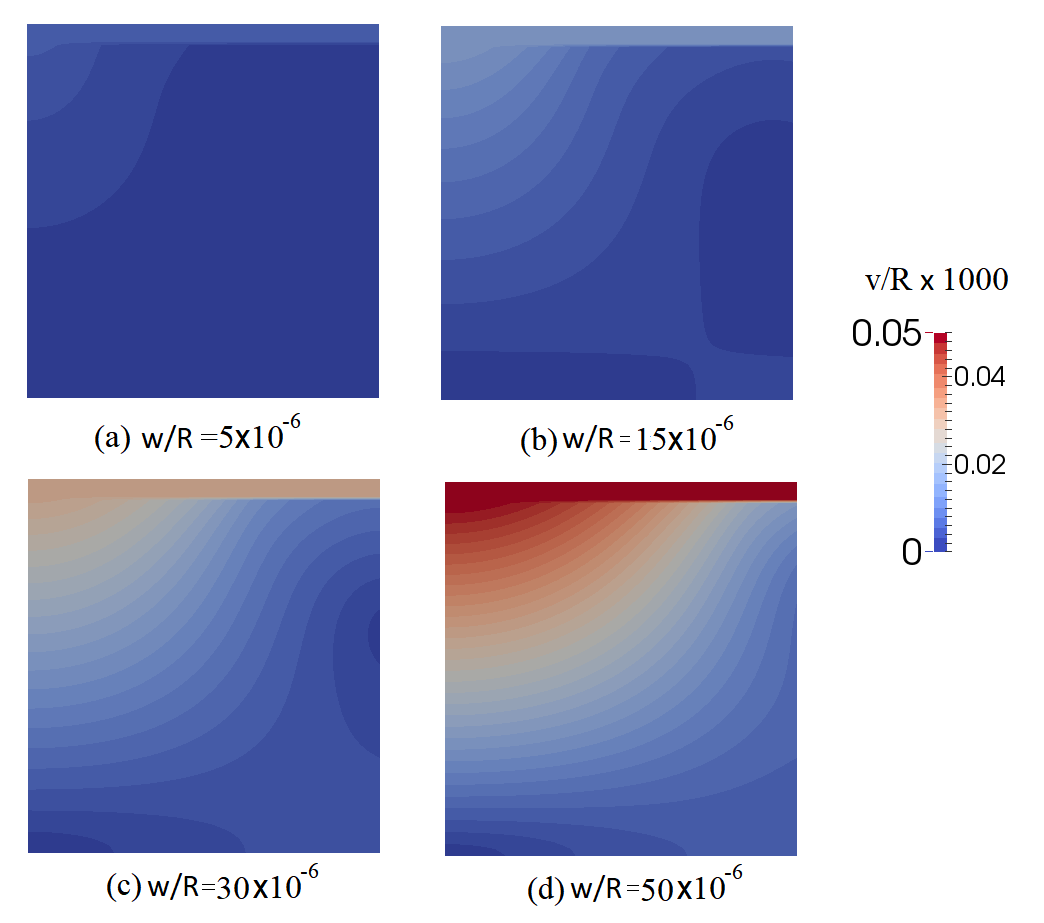}
\caption{Dimensionless vertical displacement field, $v/R$, for different imposed far-field displacements $w/R$, for $(R/x_l=100)$. $E$, $R$ and $x_l$ denote, respectively, the composite Young's modulus, the cylinder radius and the lateral size of the domain.}\label{fig8}
\end{center}
\end{figure}

\begin{figure}[h!]
\begin{center}
\includegraphics[width=.7\textwidth,angle=0]{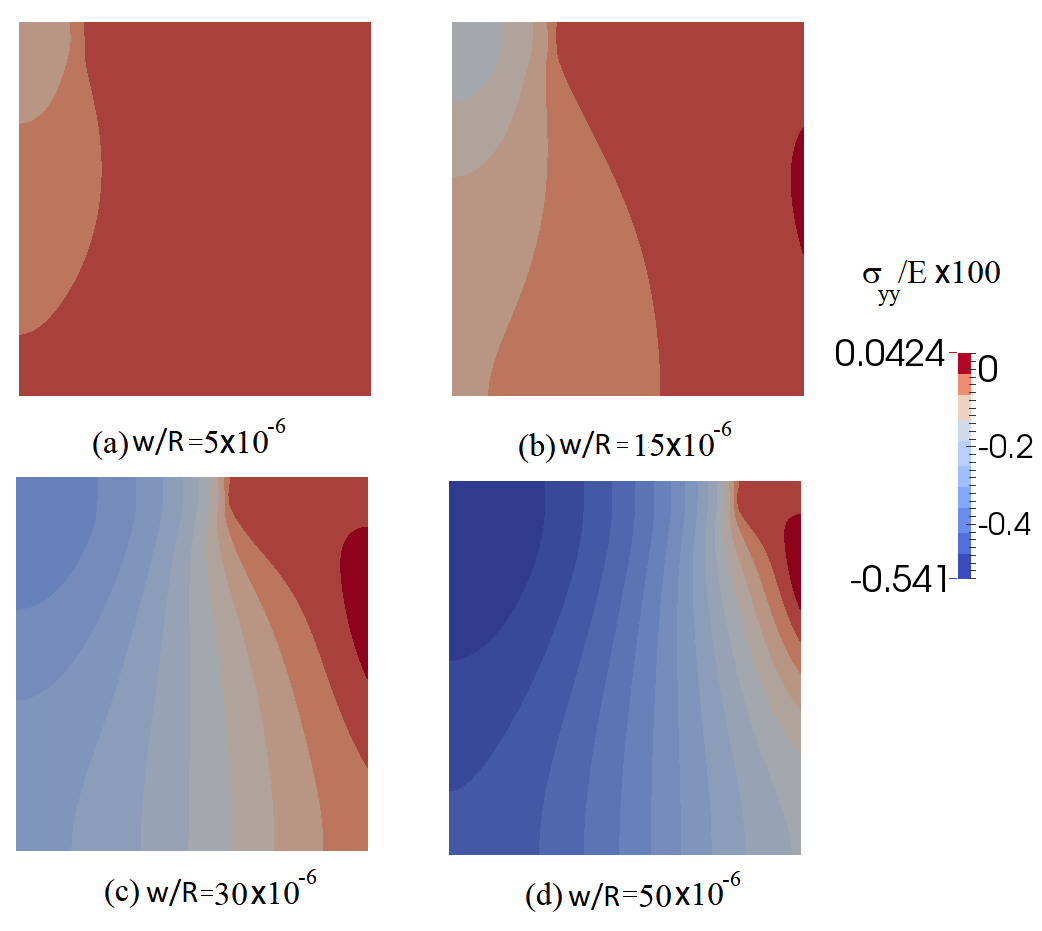}
\caption{Dimensionless vertical stress field, $\sigma_{yy}/E$, for different imposed far field displacements $w/R$, for $(R/x_l=100)$. $E$, $R$ and $x_l$ denote, respectively, the composite Young's modulus, the cylinder radius and the lateral size of the domain.}\label{fig9}
\end{center}
\end{figure}

It is also worth mentioning that the present numerical method allows the easy computation of the displacement field within the bulk, see for the case $R/x_l=100$ the contour plots of the vertical displacement field in Fig.\ref{fig8}, and also the stress field, see Fig.\ref{fig9} for the $\sigma_{yy}$ component, as a standard post-process of the finite element computations.

\subsection{Adhesive Hertzian contact problem between a cylinder and a half-plane}

The same problem analyzed in the previous section for the rigid cylinder in contact with a deformable half plane is herein examined in relation to the geometry characterized by $R/x_l=100$, considering also adhesive interactions. The analytical solution is available only for the limit case of the Johnson-Kendall-Roberts adhesion theory and it is provided in \citep{G07P1} based on the procedure described by \cite{barber_elasticity}. The authors have not found in the literature the solution corresponding to the Lennard-Jones traction law herewith employed.

In the present  analysis, the   parameters  of the Lennard-Jones formulation are set equal to  $\varepsilon=1\times 10^{-3}$ N/m, $\kappa/x_l=2.57\times 10^{-4}$, and $g_{n,0}/x_l=2.885\times 10^{-4}$, to model a relatively long-range adhesion as compared to the size of the problem, $x_l$. No cut-off distances are considered to limit adhesive interactions.

To properly capture the adhesive zone ahead of the contact strip, a finite element discretization consisting in $n_{el}=300$ finite elements along the interface is herein adopted, so that the finite element size over the cylinder radius is a small quantity: $x_l/(n_{el}R)=1/3000$. The pull-off force is computed in the present test by prescribing the following non monotonic far-field displacement path composed of two ramps throughout  the pseudo-time evolution, and  considering  that  the cylinder is in contact with the half-plane just in one point along the symmetry line which is identified as  the starting configuration. In the former ramp range (denominated as approaching ramp), the upper block (the rigid cylinder) is pushed against the half-plane by linearly increasing the imposed far-field displacement $w$ with respect to a pseudo-time variable, up to $w_{\max}/R=1\times 10^{-3}$. Then, the far-field displacement is reverted till the two surfaces are pulled apart (being denoted as separation ramp). As is well-known \citep{G07P1}, the total force during the approaching ramp is higher than the total force during the separation ramp, for the same displacement $w$, thus leading to a hysteretic energy dissipation.

\begin{figure}[h!]
\begin{center}
\includegraphics[width=.7\textwidth,angle=0]{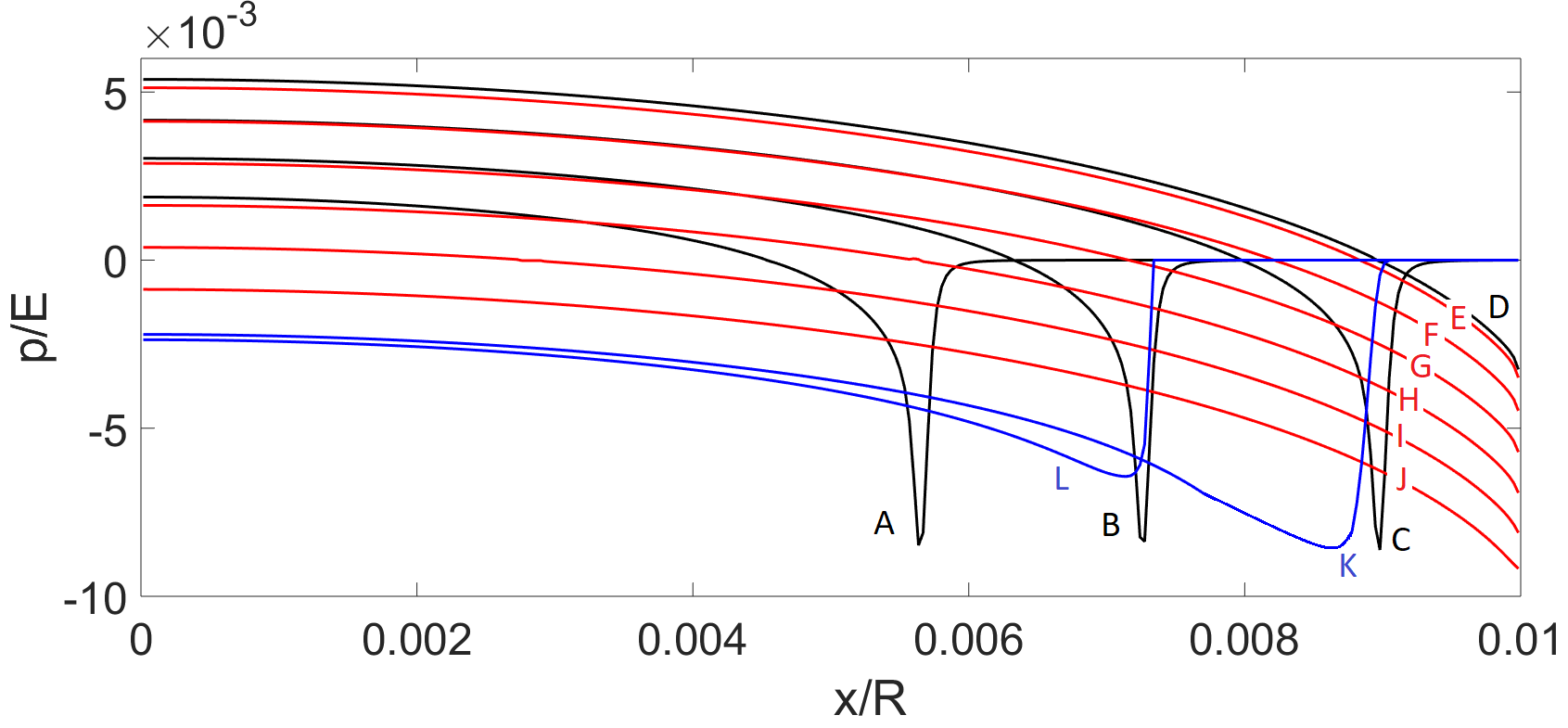}
\caption{Contact pressure $p=-p_n$ at the interface, for selected imposed far-field displacements.}\label{fig10}
\end{center}
\end{figure}

Contact pressures $p=-p_n$ are shown in Fig.\ref{fig10} for selected imposed far-field displacements. Pressures are positive valued in the contact domain $\Gamma_C^*$, while they are negative valued in the adhesive domain $\Gamma_A^*$. During the approaching phase, contact pressures have the shape of the black curves in Fig.\ref{fig10} ($w$ is increasing from A to D). Due to the action of adhesion, the contact strip becomes larger as compared to the adhesiveless case, as also expected for adhesion models with adhesive tractions constant or dependent on the normal gap, see e.g. \citep{shi}. Curves A to C display, in addition to a portion in contact and one subject to adhesive tractions, also a significant portion with negligible tractions.
The curve D corresponds to a contact configuration with a very high contact area, where the whole interface is in contact or adhesion till the boundary, without any stress-free portion.

During reverse loading, see the curves labeled from E to J in red color in Fig.\ref{fig10}, the contact area progressively shrinks and the adhesive region spreads out. This continues till the whole interface is in tension and the maximum adhesive traction is achieved at the edge of the interface. Afterwards, the interface separates from the edge, like for the propagation of an interface crack, and the adhesive zone progressively reduces, see the curve K in blue in Fig.\ref{fig10}. A further progress in the detachment leads to the curve L, which corresponds to the minimum size of the adhesive zone before the onset of an instability under displacement control. After that point, the unstable branch can only be followed by force control, as well-known in the presence of adhesion, see \citep{G07P1}.

\section{Contact between a rough profile and a half-plane}

In this section, we apply the proposed formulation to the normal contact problem between a rough profile under plain strain conditions and a half-plane, with or without adhesion.

Although the formulation does not preclude any form of roughness, that could also be provided in input as a discrete set of nodal elevations obtained by experimental measurements, we focus our attention onto the Weierstrass-Mandelbrot profile, used in \citep{Ciavarella} as a formulation to account for fractal roughness. This has been investigated in contact mechanics since it is an interesting and suitable model for fractal roughness, which encompasses a simple and closed-form  analytic equation to describe the  point-wise elevation $z(x)$:
\begin{equation}\label{WM}
z(x)=A_0\sum_{n=0}^{\infty} \gamma^{(D-2)n}\cos\left(2\pi\dfrac{\gamma^n x}{\lambda}\right),
\end{equation}
where $A_0$, $\gamma$ $(\gamma>1)$ and $D$ $(1<D<2)$ are model parameters, $\lambda$ is the longest wavelength of the profile, while $n \in \mathbb{N}$ denotes the number of length scales involved in the rough profile. The sum is usually truncated at a given $n_{\max}$, obtaining a pre-fractal rough surface.
Therefore, in the present setting, the flat boundary $\Gamma_2^*$ of the half-plane in contact with the rough profile is defined by $e_2(x)=\overline{h}_2=x_l$ and we associate roughness to $\Gamma_1^*$ by setting $e_1(x)=\overline{h}_2+h^*(x)$, where $h^*(x)=z(0)-z(x)$.

\begin{figure}[h!]
\begin{center}
\includegraphics[width=.8\textwidth,angle=0]{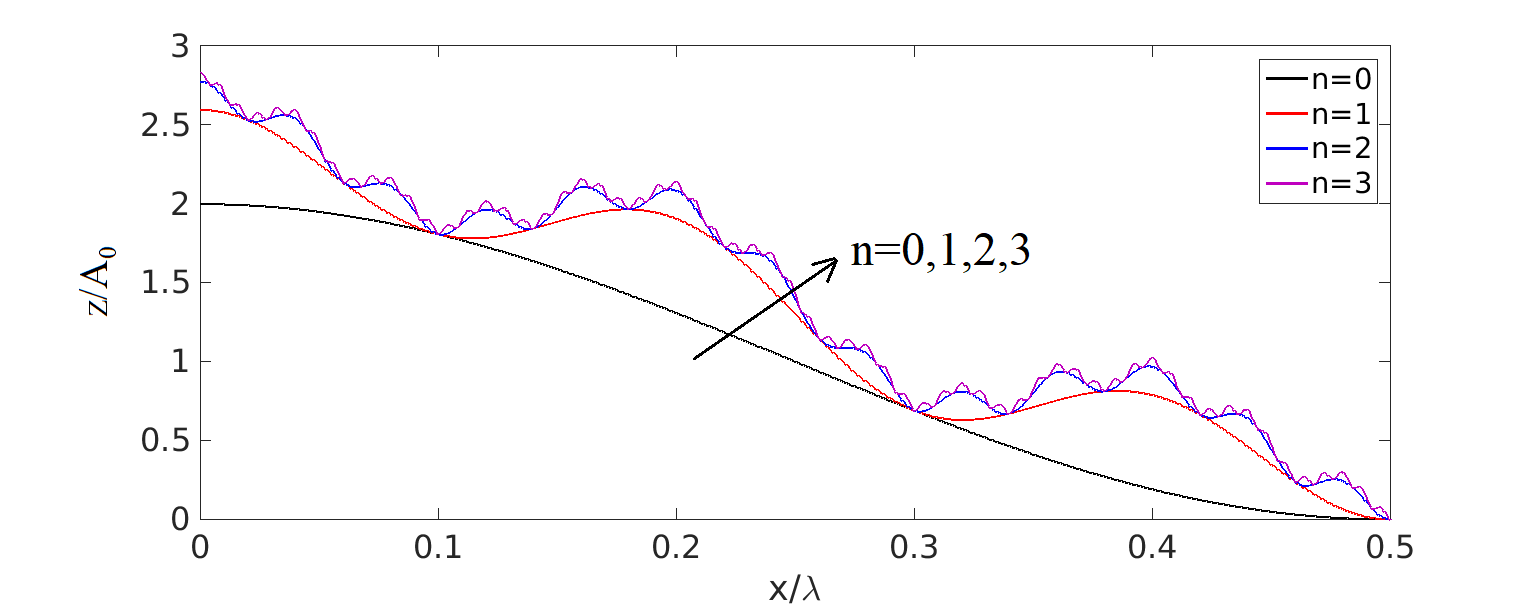}
\caption{Weierstrass-Mandelbrot profiles depending on the resolution parameter $n$ from 0 up to 3, fractal dimension $D=1.25$, and scaling parameter $\gamma=5$.}\label{fig11}
\end{center}
\end{figure}

For $n=0$, a single co-sinusoidal function of $x/\lambda$, with a periodicity of $\lambda$ in the horizontal direction, is obtained. This function has a peak (maximum) at $x/\lambda=0$ and at $x/\lambda=1$, and a valley (minimum) at $x/\lambda=0.5$ (see the black curve shown in Fig.\ref{fig11} in the range $0\le x/\lambda\le 0.5$, symmetric in the range $0.5\le x/\lambda\le 1$). The profile with $n=1$ is obtained by superimposing to the profile with $n=0$ a finer co-sinusoidal function with a shorter wavelength $\lambda/\gamma$ and a re-scaled amplitude $A_0\gamma^{(D-2)}$, see the red curve in Fig.\ref{fig11}. The addition of further length scales proceeds in a similar manner by adding finer roughness with shorter wavelengths up to $\lambda/\gamma^n$ and re-scaled amplitudes $A_0\gamma^{(D-2)n}$, which leads to the profiles with $n=2$ and $n=3$ shown in Fig.\ref{fig12}, depicted in blue and purple, respectively.

In the sequel, we investigate three issues involving roughness by exploiting the features of the proposed variational approach and its versatility for the quantitative investigation of contact problems where analytical solutions are difficult to be derived: $(i)$ the effect of length scales of roughness on the adhesiveless normal contact problem, quantitatively reproducing the trends on the lacunarity of the contact problem qualitatively predicted by the semi-analytical approach in \citep{Ciavarella} which is exact only in the case of full contact; $(ii)$ the assessment of the effect of having a finite size domain $\Omega_2$ instead of a half-plane on the contact response, which is difficult to be handled analytically and also using numerical methods like BEM that require a modification of the Green function; $(iii)$ the investigation of the problem of adhesive contact, whose analytical solution is not available due to the non-compactness of the contact domain and it is a challenging problem also for the existing BEM and FEM numerical strategies.

\subsection{The effect of length scales of roughness on the adhesiveless contact problem}

The frictionless and adhesiveless normal contact problem between a rigid indenter with the Weierstrass-Mandelbrot profile described by Eq.\eqref{WM} and a deformable half-plane has been thoroughly investigated in \citep{Ciavarella}. The complexity arises by the fact that the analytical solution of the problem for a single co-sinusoid $(n=0)$ can be deduced by the application of the Westergaard method only in the case of full contact conditions, where the deformed shape of the half-plane boundary is exactly given by the rigid profile of the indenter. Partial contact is much more difficult to be analytically assessed and the authors in \citep{Ciavarella} proposed an approximate recursive analytical method inspired by the Archard contact model \citep{CD00} to estimate the contact pressure distribution at a scale of roughness $n$ from the solution obtained at the coarser scale $n-1$. Further insight into the real contact area led to the key conclusion that, for a given contact pressure, the real contact area is steadily decreasing with $n$, tending to zero to the fractal limit of $n_{el}\to \infty$, proving the so-called \emph{lacunarity} of the contact domain. In \citep{ciava}, the incremental normal contact stiffness was investigated, and it was found that it is mostly governed by the long wavelength components of the profile, in agreement with theoretical arguments \citep{Barber03}, see also a recent study for fractal rough surfaces in \citep{PB11}.

To solve the present problem numerically, we take into account the $\lambda$-periodicity of the profile by imposing periodic boundary conditions at $x/\lambda=0$ and at $x/\lambda=1$. Then, due to the symmetry of the geometry and loading, which is an imposed far-field displacement $w$ to the upper edge of the domain $\Gamma_1$ as in the previous examples, the contact predictions are shown only in the range $0\le x/\lambda\le 0.5$, since the solution is symmetric in the range $0	.5\le x/\lambda\le 1$. The finite element discretization is chosen to have $h/\lambda_{\min}=0.125$, where $\lambda_{\min}=\lambda/\gamma^n$, in order to properly resolve the contact traction distribution for any rough profile.

\begin{figure}[h!]
\begin{center}
\subfigure[$n=0$]{\includegraphics[width=.7\textwidth,angle=0]{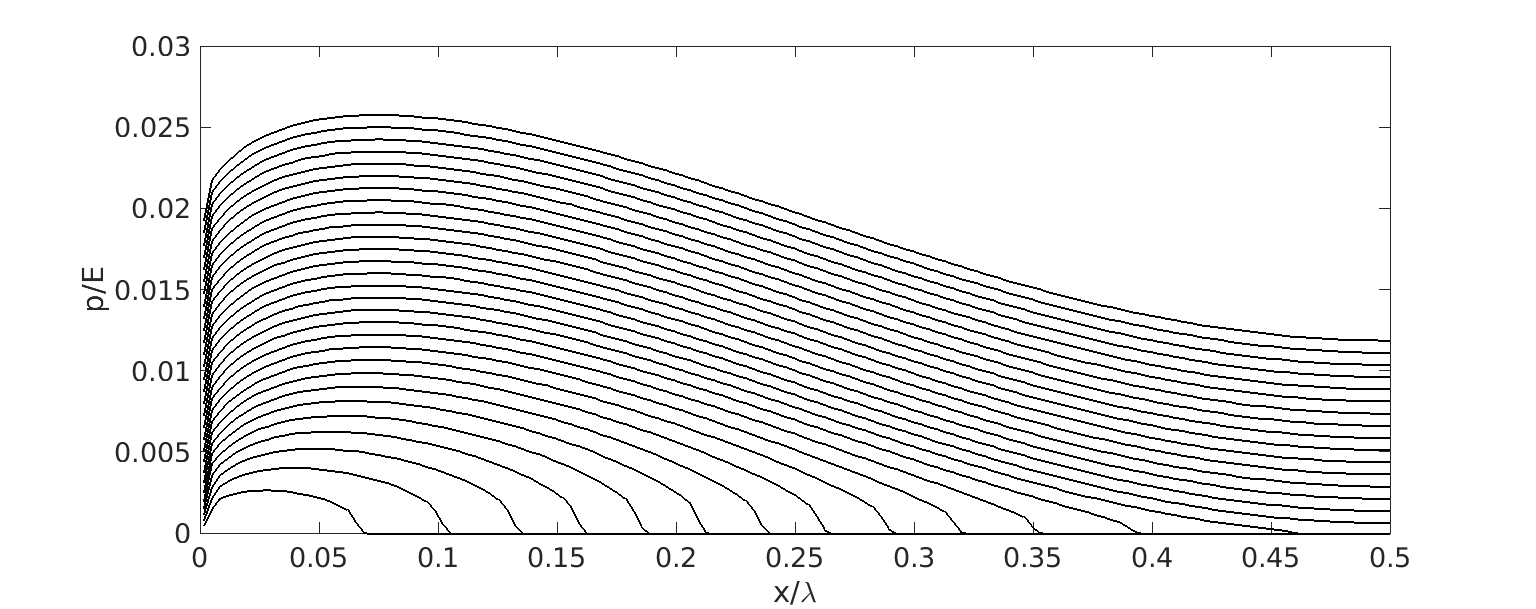}\label{fig12a}}\\
\subfigure[$n=1$]{\includegraphics[width=.7\textwidth,angle=0]{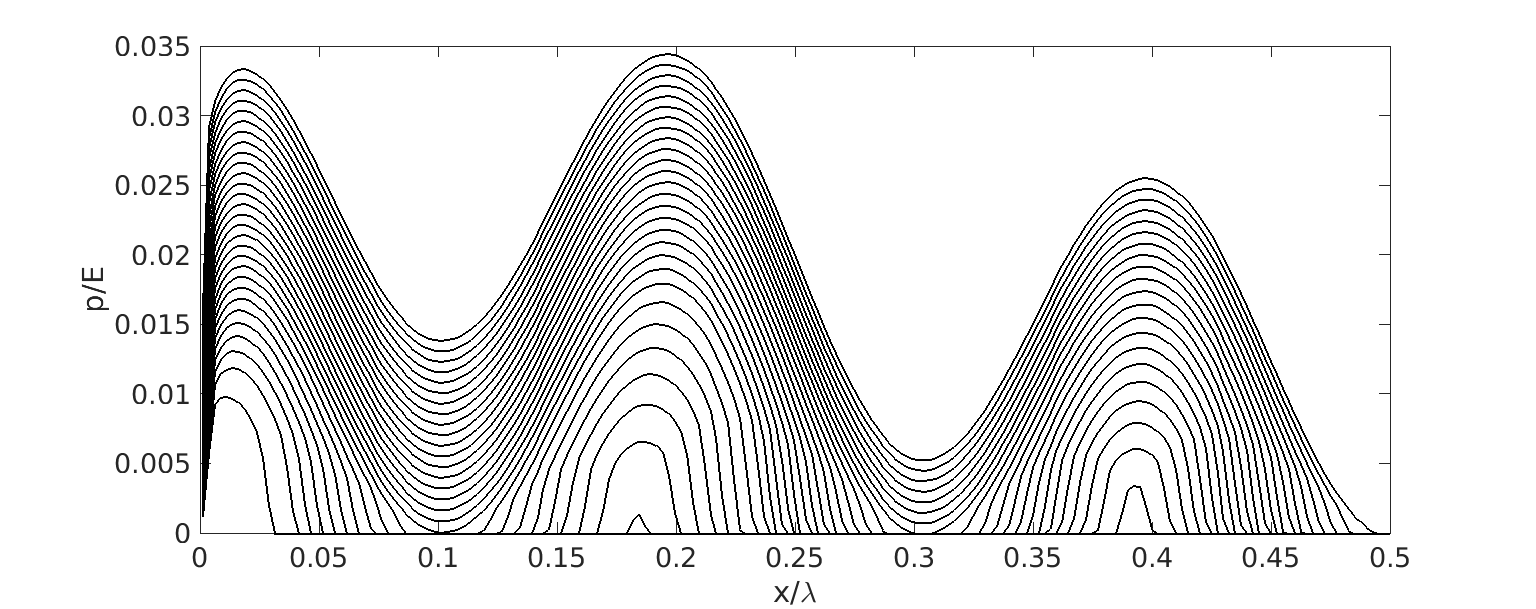}\label{fig12b}}\\
\subfigure[$n=2$]{\includegraphics[width=.7\textwidth,angle=0]{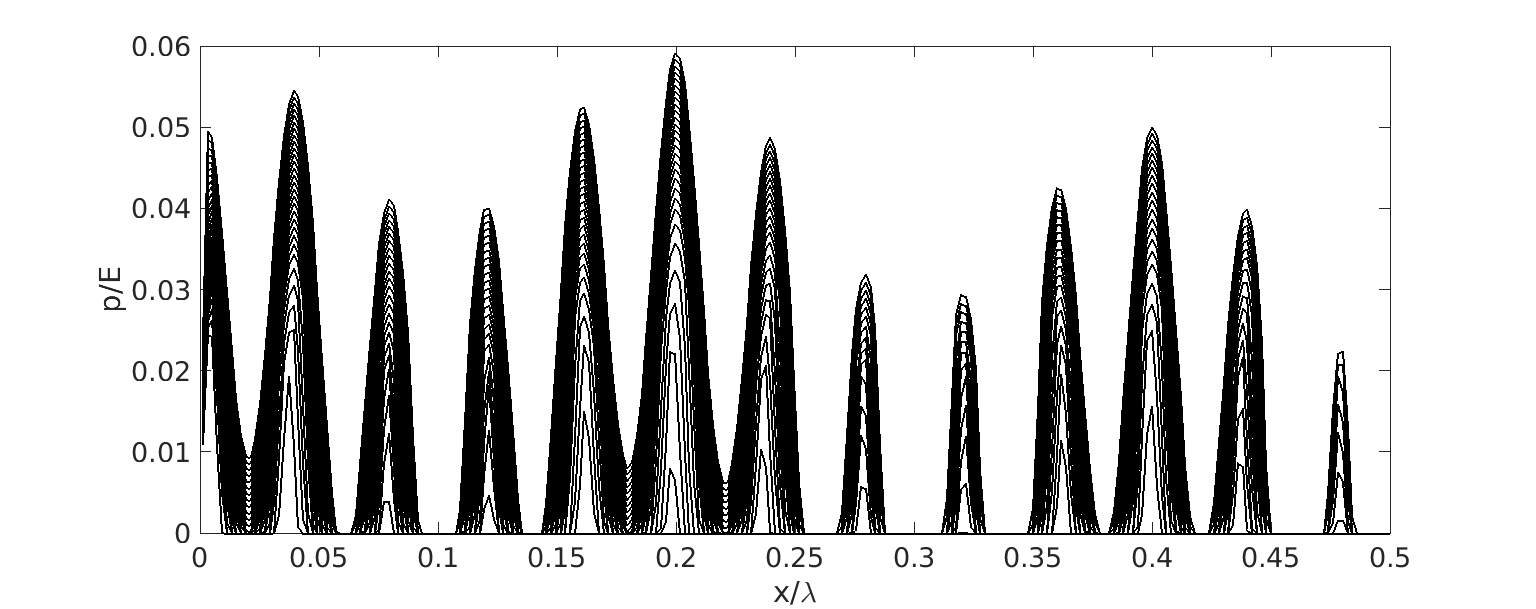}\label{fig12c}}\\
\subfigure[$n=3$]{\includegraphics[width=.7\textwidth,angle=0]{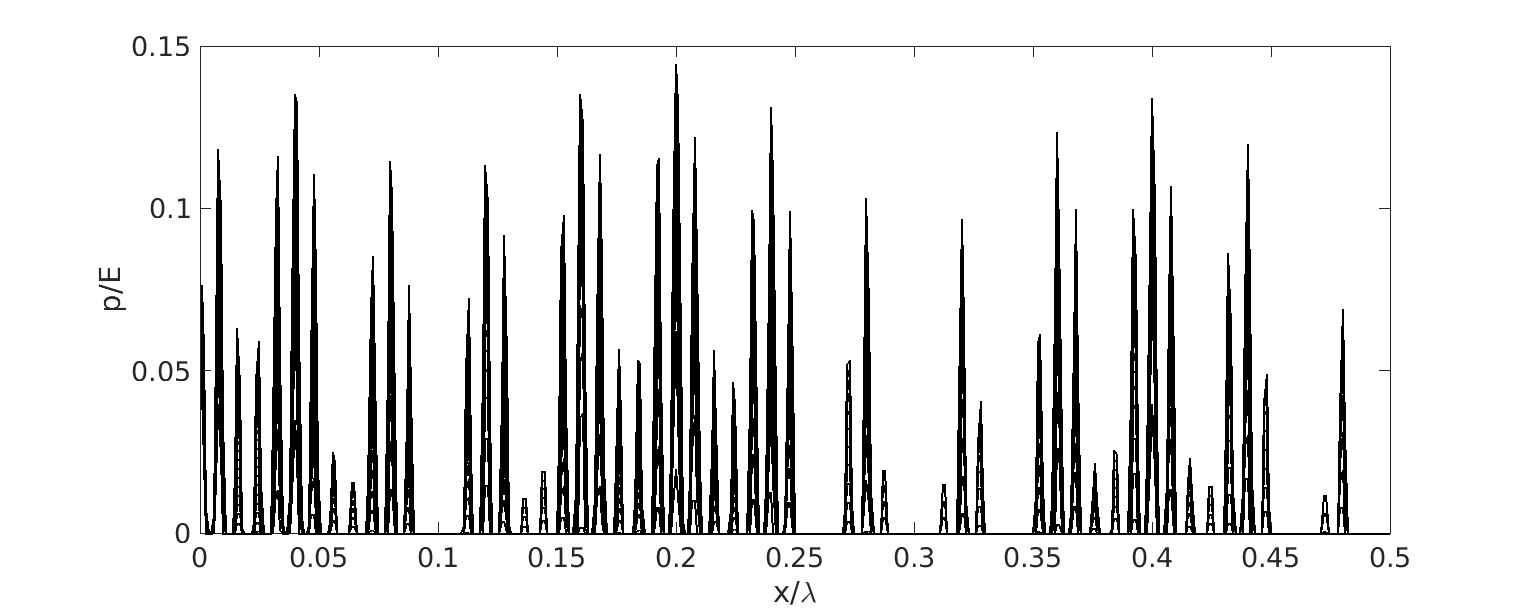}\label{fig12d}}
\caption{Dimensionless contact pressure along the interface depending on the resolution parameter $n$, $D=1.25$, $\gamma=5$. $E$, $\lambda$ denote, respectively, the composite Young's modulus and the longest wavelength of the profile.}\label{fig12}
\end{center}
\end{figure}

\begin{figure}[h!]
\begin{center}
\includegraphics[width=.7\textwidth,angle=0]{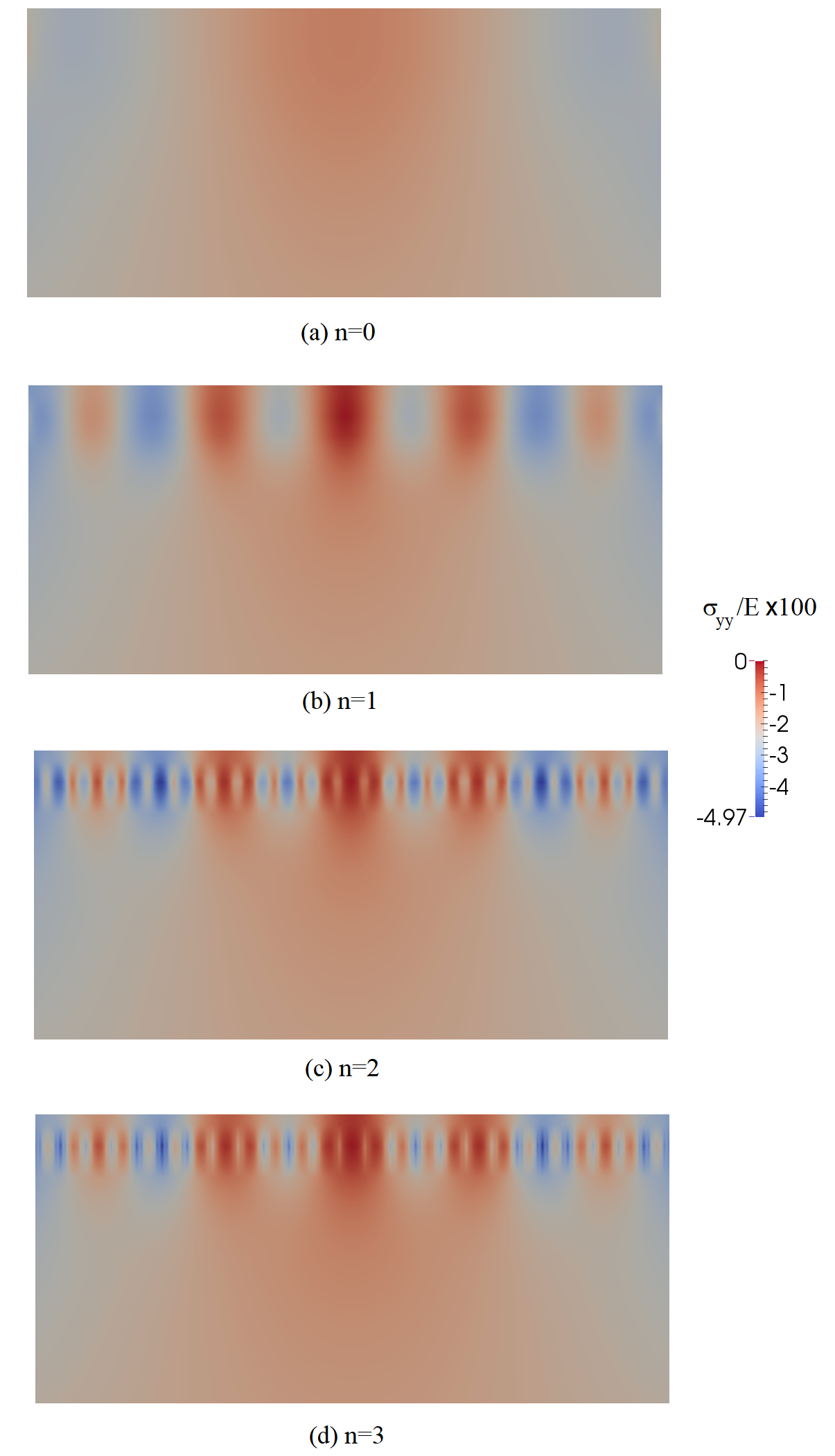}
\caption{Dimensionless vertical contact stress $\sigma_{yy}$ in the bulk depending on the resolution parameter $n$, $D=1.25$, $\gamma=5$.}\label{fig13}
\end{center}
\end{figure}

Contact pressures $p=-p_n$ along the interface are shown in Fig.\ref{fig12} for Weierstrass-Mandelbrot profiles with $D=1.25$, $\gamma=5$, $A_0/\lambda=0.0025$, and different values of the resolution parameter $n$. The set of curves shown in the subfigures correspond to the same values of the imposed far-field displacements. The solution for $n=0$ shows that the contact domain is a continuous increasing function of $w$ and its spreads all over the interface, achieving the full contact condition. A refinement of the profile by adding another term in the series $(n=1)$ leads to an increase in the contact pressures for the same imposed displacement $w$. The full contact condition is also obtained, but it is clearly delayed. The addition of further length scales ($n=2$ and $n=3$) drastically increases the value of the contact pressures and it reduces the real contact area which localizes near the asperities (the local maxima of the rough profile), inhibiting the achievement of full contact conditions. The contour plot of the dimensionless stress field component $\sigma_{yy}/E$ in the bulk in correspondence to the maximum imposed displacement is shown in Fig.\ref{fig13}, quantitatively showing up to which depth from the interface the non-uniformity in the contact tractions influences the stress field component $\sigma_{yy}$ before smearing out its effect.

\begin{figure}[h!]
\begin{center}
\includegraphics[width=.99\textwidth,angle=0]{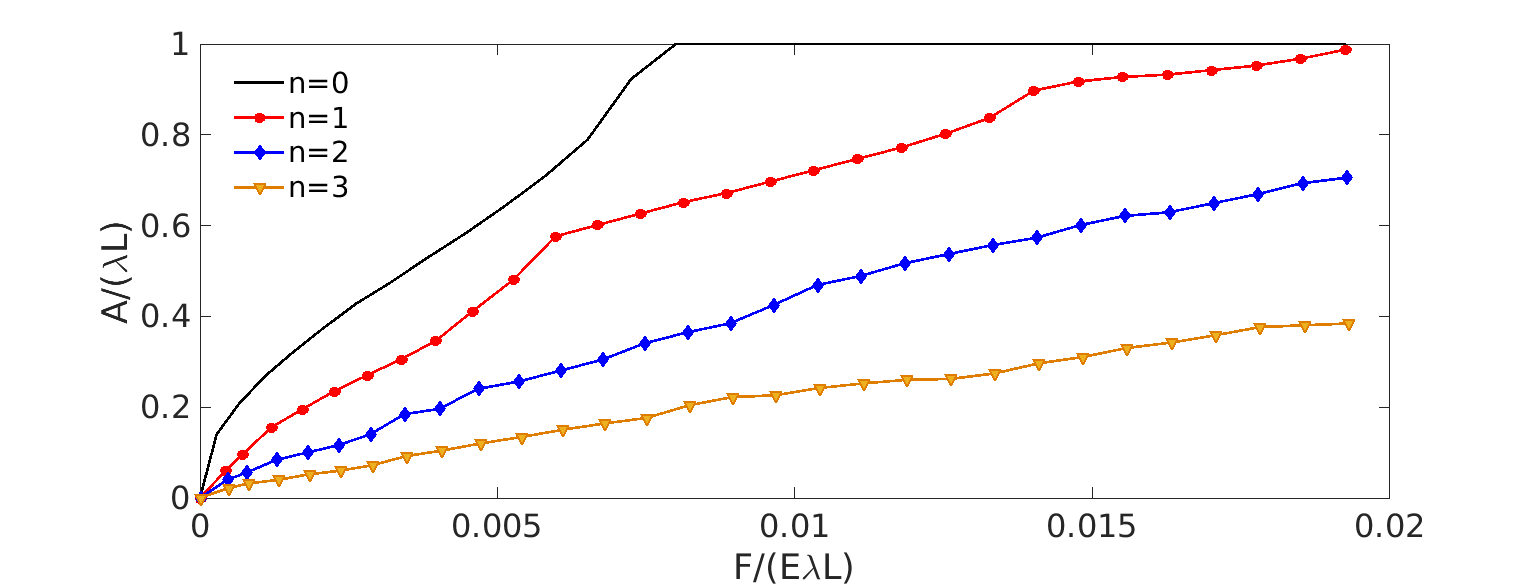}
\caption{Dimensionless real contact area vs. dimensionless contact force depending on the resolution parameter $n$, $D=1.25$, $\gamma=5$.}\label{fig14}
\end{center}
\end{figure}

A quantitative assessment of the lacunarity effect can be assessed from the dimensionless real contact area, $A/(\lambda L)$, vs. total compressive normal force, $F/(E\lambda L)$, where $L$ denotes the out-of-plane thickness of the model that in the present plane strain solution is equal to unity (see Fig.\ref{fig14}).
In this plot, the value $A/(\lambda L)=1$ corresponds to full contact. The addition of finer length scales of roughness leads to a progressive reduction of the real contact area in correspondence of the same applied force, with a trend consistent with analytical estimates in \citep{Ciavarella} strictly valid for very large values of $n$.
Regarding the incremental normal contact stiffness, we do not report results here, but we observed a appreciable difference between the cases characterized by $n=0$ and $n=1$, while further refinement of roughness ($n=2$ and $n=3$) introduced only negligible perturbations. This is also consistent with theoretical arguments in \citep{Barber03}, where it was proved that waviness plays a dominant role over fine scale roughness as far as the incremental normal contact stiffness is concerned.

\subsection{The effect of the domain size in the adhesiveless contact problem}

The frictionless contact problems without adhesion investigated in the previous sections were conducted by considering a square deformable domain of lateral size $\lambda$. The vertical size $t=\lambda$ was sufficiently large to be representative of half-plane contact problems. The study of the effect of reducing  $t$ is indeed of interest in many applications, and cannot be easily solved using BEM unless modified Green's functions are determined. This finite-size issue, as well as the problem of having material inhomogeneities like voids or inclusions inside the domain $\Omega_2$, can be instead easily tackled by the current FEM-based framework.

\begin{figure}[h!]
\begin{center}
\includegraphics[width=.99\textwidth,angle=0]{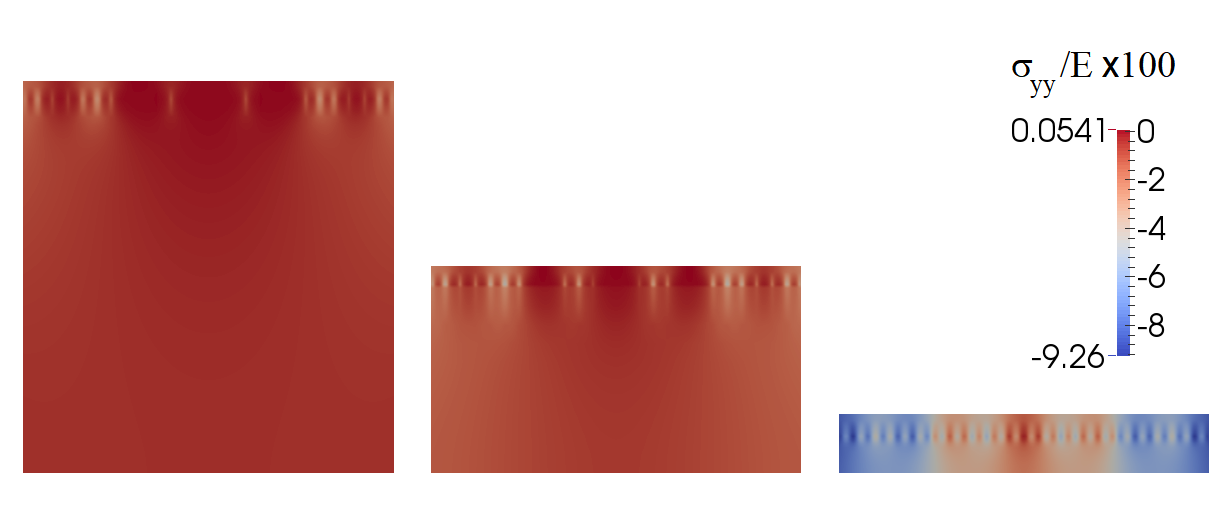}
\caption{Dimensionless vertical contact stress $\sigma_{yy}$ in the bulk depending on the domain size, $t$, for a far-field imposed displacement $\Delta/g_0=3$ ($D=1.25$, $\gamma=5$, $n=2$).}\label{fig15}
\end{center}
\end{figure}

\begin{figure}[h!]
\begin{center}
\includegraphics[width=.99\textwidth,angle=0]{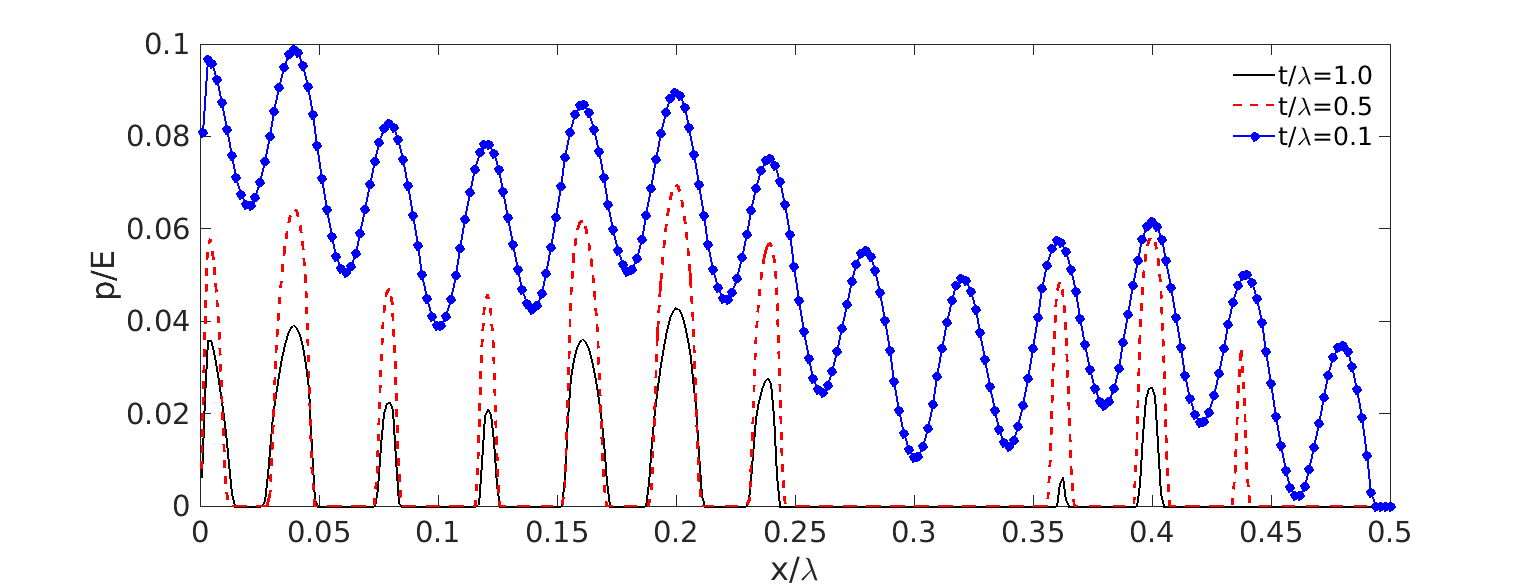}
\caption{Dimensionless contact pressure along the interface depending on the domain size, $t$, for a far-field imposed displacement $\Delta/g_0=3$ ($D=1.25$, $\gamma=5$, $n=2$). $E$ and $\lambda$ denote, respectively, the composite Young's modulus and the longest wavelength of the profile.}\label{fig16}
\end{center}
\end{figure}

In this section, we replicate  the contact mechanics simulations for the Weierstrass-Mandelbrot rough profile analyzed in the previous section in reference to the case with parameters $D=1.25$, $\gamma=5$, $A_0/\lambda=0.0025$, $n=2$. In particular, we compare the solution for an imposed far-field displacement $w/A_0=3$, for $t/\lambda=1$, 0.5, and 0.1. The contour plots of the corresponding dimensionless stress field component $\sigma_{yy}/E$ is shown in Fig.\ref{fig15} for the three cases, highlighting the increase in the stress level for the same amount of imposed displacement $w$ as a result of the increased stiffness of the finite-size domain $\Omega_2$. This is also evident from the comparison in terms of contact tractions along the interface in Fig.\ref{fig16}, which shows that thin substrates are very likely to experience full contact conditions as compared to the half-plane geometry, even in the presence of roughness that is usually promoting lacunarity of the contact domain.

\subsection{Adhesive contact problem with roughness}

The frictionless normal contact problem between a rigid Weierstrass-Mandelbrot plane strain indenter and a deformable half-plane analyzed in Sec. 6 is herein investigated by including also adhesion into the picture. Regarding the  elastic and the adhesion model parameters, the reader is referred to the input data listed in Sec. 5.2. The parameters for the Weierstrass-Mandelbrot profile are also the same as those in Sec. 6.1. Similarly, two ramps for the imposed far-field displacements are herein considered: $(i)$ an approaching stage, characterized by an increasing $w$ to put the indenter in contact with the half-plane up to $w/A_0=3$, and $(ii)$ a separation stage, where $w$ is reduced until complete separation of profile is reached. As compared to smooth profiles, the modeling complexity here regards the non-compactness of the contact domain, which imposes severe limitations in the use of semi-analytical methods. For instance, the method proposed in \citep{G07P1} strictly applies to profiles whose shapes are leading to monotonically increasing gaps to avoid partial contact, which is a phenomenon occurring in the present case.

We first focus our attention on the analysis of the evolution of the pressure distribution at the interface, $p/E$, and of the corresponding normal gap $g_n/A_0$ for a Weierstrass-Mandelbrot profile with roughness parameter $n=1$.

\begin{figure}[h!]
\begin{center}
\subfigure[Interface traction]{\includegraphics[width=.7\textwidth,angle=0]{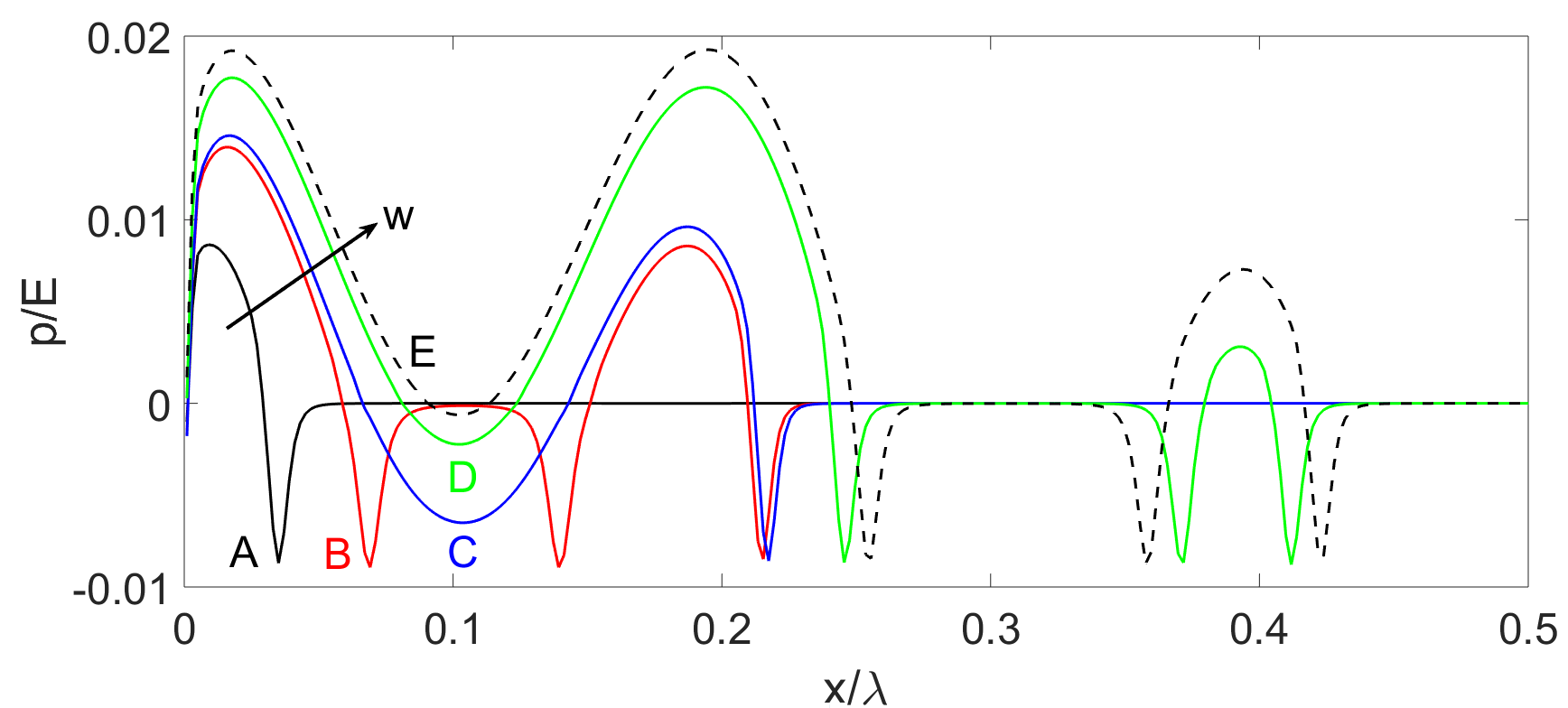}\label{fig17a}}\\
\subfigure[Normal gap]{\includegraphics[width=.7\textwidth,angle=0]{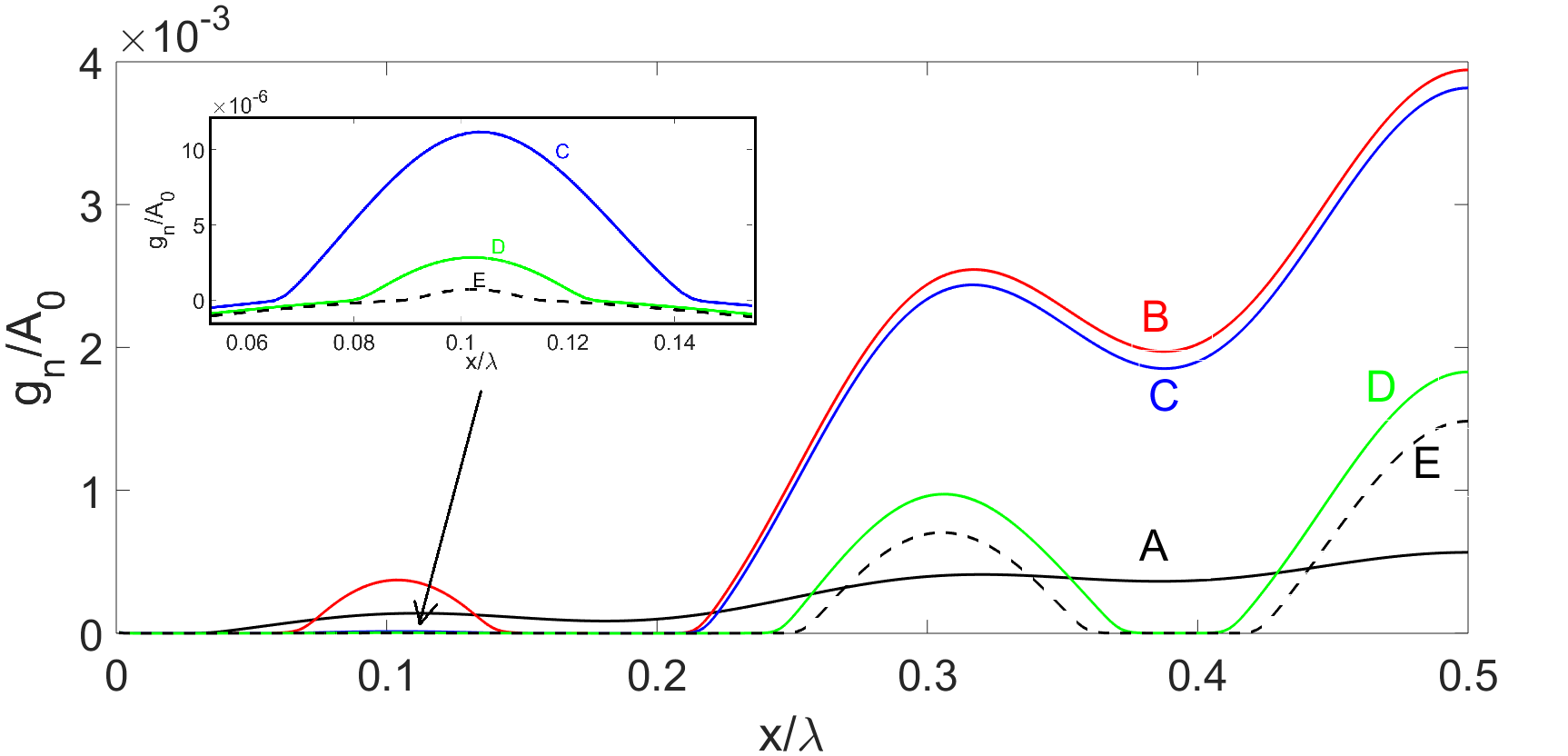}\label{fig17b}}
\caption{Dimensionless contact pressure along the interface for different far-field imposed displacements $w$ during the approaching stage (Weierstrass-Mandelbrot profile with $A_0/\lambda=0.0025$, $D=1.25$, $\gamma=5$, $n=1$). $E$ and $\lambda$ denote, respectively, the composite Young's modulus and the longest wavelength of the profile.}\label{fig17}
\end{center}
\end{figure}

\begin{figure}[h!]
\begin{center}
\subfigure[Interface traction]{\includegraphics[width=.7\textwidth,angle=0]{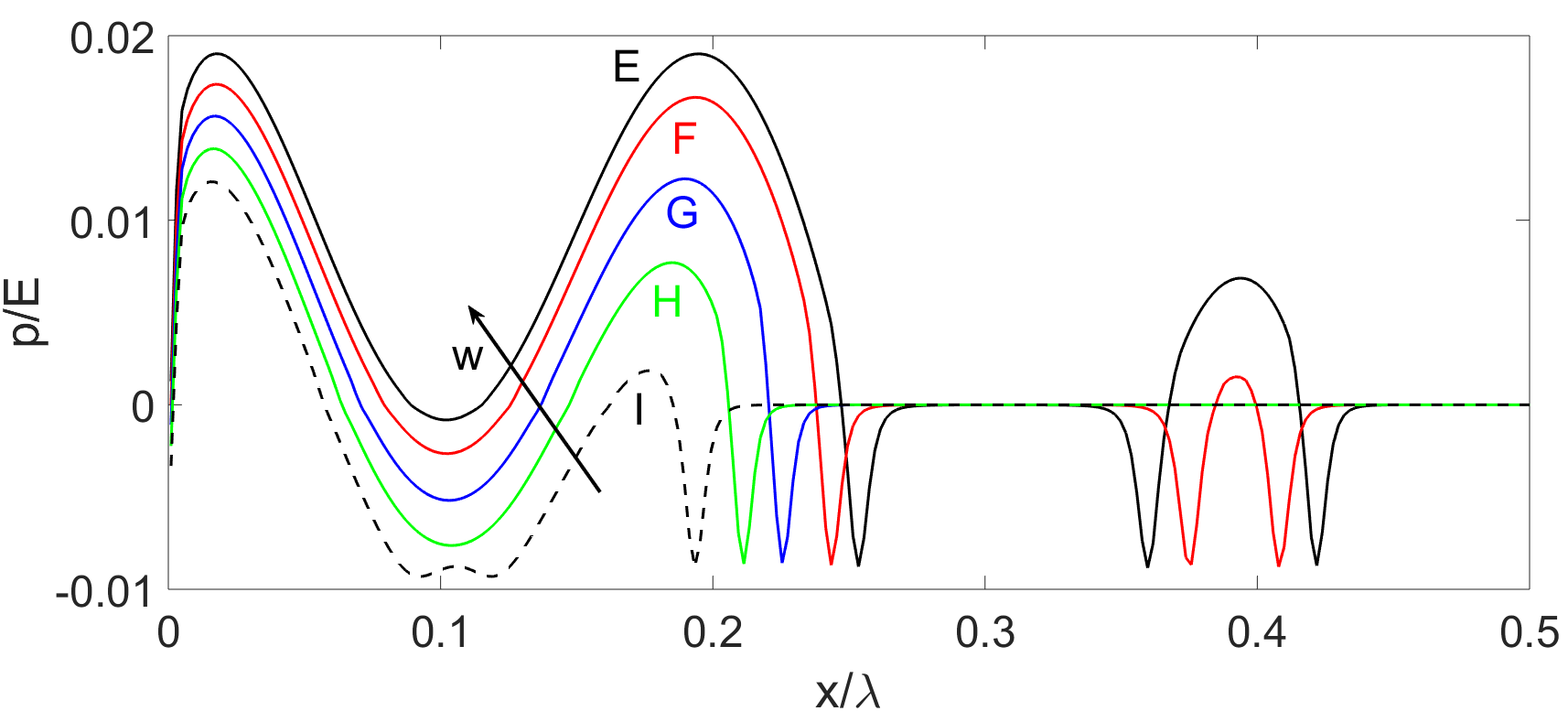}\label{fig18a}}\\
\subfigure[Normal gap]{\includegraphics[width=.7\textwidth,angle=0]{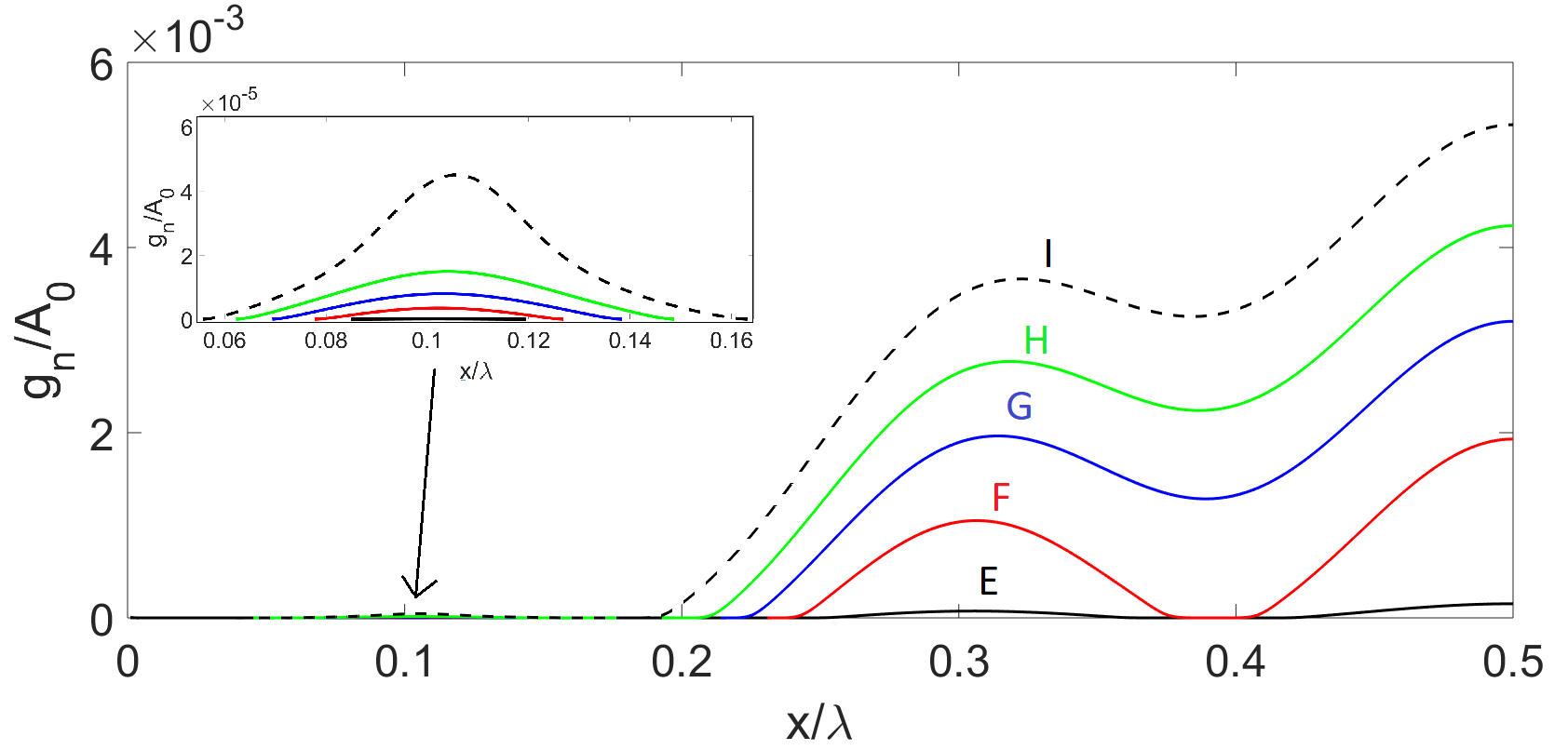}\label{fig18b}}
\caption{Dimensionless contact pressure along the interface for different far-field imposed displacements $w$ during the separation stage (Weierstrass-Mandelbrot profile with $A_0/\lambda=0.0025$, $D=1.25$, $\gamma=5$, $n=1$). $E$ and $\lambda$ denote, respectively, the composite Young's modulus and the longest wavelength of the profile.}\label{fig18}
\end{center}
\end{figure}

Numerical predictions for selected values of $w$ are shown in Fig.\ref {fig17} for the approaching stage, and in Fig.\ref{fig18} for the separation stage. During the approach, contact takes place near the first peak at $x/\lambda=0$, with contact pressures (positive valued in the plot) followed by adhesive tractions (negative valued) significant within a small distance after the end of the contact strip (black solid curve A in Fig.\ref{fig17a}). A further increase in the closing displacement $w$ leads to an increase in the contact and adhesive domains in a self-similar manner, leading to curves intermediate from the curve A and the curve B. Within that evolution, another peak comes into contact near $x/\lambda\cong 0.17$ (see the red curve B which highlights the presence of another contact strip along the rough profile).
A further increase in $w$ leads to the curve C which now looks quite different from the previous traction distributions in the region between the two peaks near $x/\lambda\cong 0.1$. The normal gap is significantly diminished in that region, as it can be noticed from the magnification in Fig.\ref{fig17b}, and the adhesive zones of the two peaks in between the two contact strips start interacting. Such an interaction leads to adhesive tractions in the ascending branch of the adhesive law based on the Lennard-Jones potential in between $g_n=0$ and $g_n$ corresponding to the maximum adhesive traction. As a result, the corresponding adhesive traction distribution along the interface near $x/\lambda\cong 0.1$ presents now a bell-shaped form, with a fully pressurized gap. The size of such pressurized gap is then diminished by further increasing $w$ (curve D) due to the growth of the contact strips. We also observe that another peak comes in contact near $x/\lambda\cong 0.38$. This proceeds up to curve $E$, which corresponds to the maximum imposed far-field displacement $w=3A_0$.

The unloading stage, whose numerical predictions are shown in Fig.\ref{fig18}, starts from the curve E and proceeds, by progressively reducing $w$, to the sequence of curves F, G, H, and I, just before the complete separation of the entire profile. During the separation stage, the contact strips shrink and the adhesive ones increase.
However, it is remarkable to note that load reversal is leading to pressures that do not follow the distributions experienced during the approaching stage. As a result, hysteretic energy dissipation takes place due to the non coincident loading and unloading paths, as well known in adhesive contact problems.

\begin{figure}[h!]
\begin{center}
\includegraphics[width=.99\textwidth,angle=0]{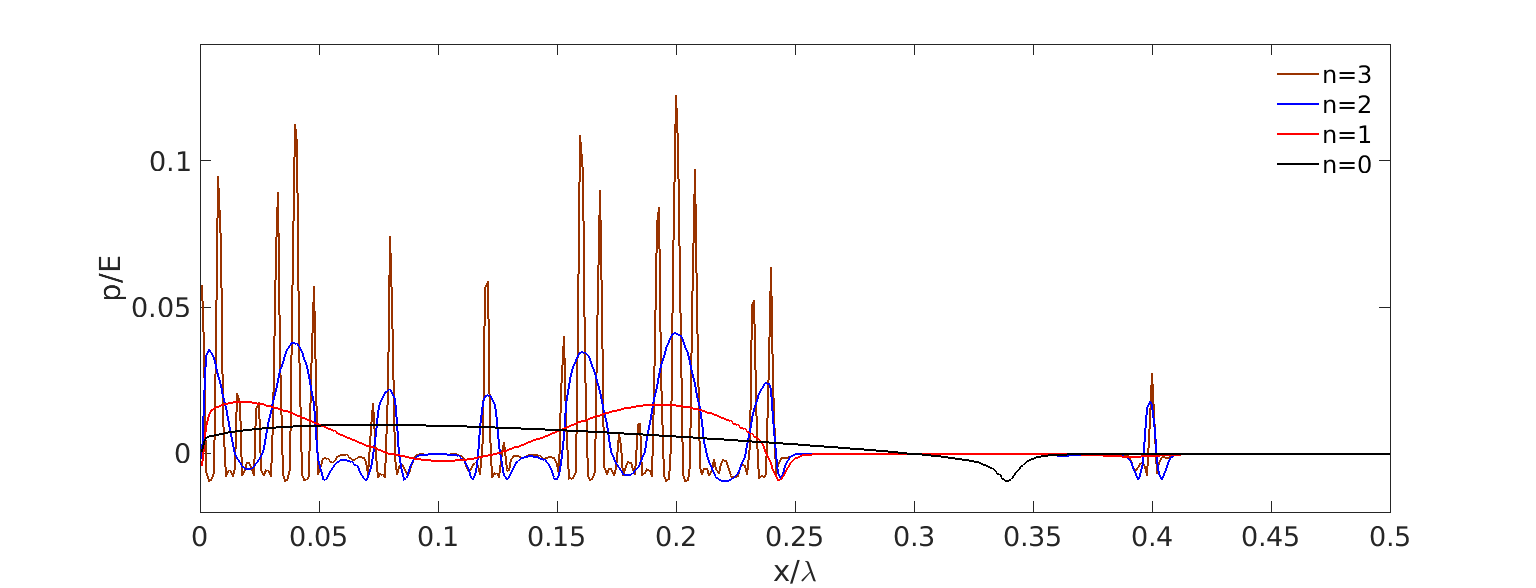}
\caption{Dimensionless contact pressure along the interface for different values of the resolution parameter $n$, for the same far-field imposed displacement $w/A_0=2.4$ (Weiestrass-Mandelbrot surfaces with $A_0/\lambda=0.0025$, $D=1.25$, $\gamma=5$). $E$ and $\lambda$ denote, respectively, the composite Young's modulus and the longest wavelength of the profile.}\label{fig19}
\end{center}
\end{figure}

Once the fundamental mechanisms taking place during adhesive contact in the presence of roughness have been now elucidated, the role played by the length scales of roughness is finally discussed by comparing the pressure distributions for the same far-field imposed displacement $w/A_0=2.4$ but for Weierstrass-Mandelbrot profiles with variable resolution parameter $n$ from 0 up to 3, see Fig.\ref{fig19}.

Numerical predictions show that the addition of length scales of roughness by increasing $n$ leads to an increase in the maximum tractions in the non-compact contact regions, as also noticed for the adhesiveless scenario. On the other hand, the adhesive contact pressure is always bounded by the fact that the adhesive constitutive relation has a maximum adhesive traction that cannot be overcome.

The key effect of roughness can be well appreciated by comparing tractions for $n=0$ and the other predictions for larger $n$. For $n=0$ (black curve), a single compact contact domain occurs, with a corresponding zone of the interface ahead of the edge of the contact strip with adhesive tractions rapidly decaying in space. For $n=1$ (red curve), two contact strips occur and the contact domain becomes non-compact. As a result, a fully pressurized region between the two peaks in contact takes place. The same trend, even more pronounced, can be observed by a further increase in $n$, which leads to very distributed non-compact contact strips along the profile, and many fully pressurized adhesive gaps in between the peaks in contact.

\section{Discussion and conclusion}

The proposed variational approach and its related finite element formulation overcome the major difficulties of standard finite element techniques in dealing with non-planar or even rough boundaries. So far, the research community has attempted to resolve such drawback by explicitly modeling waviness using high-order interpolation schemes requiring ad hoc finite element discretization techniques (Bezier interpolation, NURBS finite elements, or adaptive mesh refinement schemes, see e.g. \citep{HPMR,wriggers,laura,oysu}). Albeit quite promising for smooth contact problems, few attempts have been made to tackle multi-scale roughness with such techniques \citep{HPMR}.

In this study, we followed a completely different path, treating the interface as nominally flat, and embedding the non-planarity and roughness in the formulation as a correction to the normal gap.
This strategy allows using a linear finite element interpolation without compromising the accuracy. The use of low order interpolation schemes for the interface allows, as a direct consequence, the use of the same interpolation also for the continuum, with a significant reduction of nodal degrees of freedom as compared to higher order interpolation.
As a result, the present formulation opens new perspectives for the use of finite elements in contact problems with rough boundaries, as an alternative strategy to the boundary element method.

The examples addressed in this study have therefore highlighted the capability of the method to pass benchmark tests regarding Hertzian contact problems that are also typically used to assess the accuracy of higher order finite element schemes, showing an excellent performance in spite of the relatively coarse discretization used. Then, the selection of contact problems involving roughness have highlighted the potentiality of the method in solving problems difficult to be addressed by analytical or by the boundary element method, due to the finite size of the problem, the non compactness of the contact domain,  the simultaneous presence of strongly nonlinear contact and adhesion phenomena, and the analysis of non-monotonous loading paths.

Further steps left for investigation should regard the application of the method to three-dimensional topologies, the study of the interplay with plasticity in the bulk, and the study of frictional contact problems.

\section*{References}

\bibliographystyle{model2-names}\biboptions{authoryear}


\end{document}